\documentclass[aps,prl,twocolumn,superscriptaddress]{revtex4-2}

\usepackage{physics}
\usepackage{graphicx}
\usepackage{xcolor}
\usepackage{blindtext}
\usepackage{url} 
\usepackage{soul}

\begin{document}


\title{Optimization and readout-noise analysis of a warm vapor EIT memory on the Cs D1 line}


\author{Luisa Esguerra}
\email[]{luisa.esguerrarodriguez@dlr.de}
\affiliation{German Aerospace Center (DLR), Institute of Optical Sensor Systems, Rutherfordstr. 2, 12489 Berlin, Germany}
\affiliation{TU Berlin, Institute for Optics and Atomic Physics, Hardenbergstr. 36, 10623 Berlin, Germany}

\author{Leon Me\ss ner}
\affiliation{German Aerospace Center (DLR), Institute of Optical Sensor Systems, Rutherfordstr. 2, 12489 Berlin, Germany}
\affiliation{Institut für Physik, Humboldt-Universit\"at zu Berlin, Newtonstr. 15, 12489 Berlin, Germany}

\author{Elizabeth Robertson}
\affiliation{German Aerospace Center (DLR), Institute of Optical Sensor Systems, Rutherfordstr. 2, 12489 Berlin, Germany}
\affiliation{TU Berlin, Institute for Optics and Atomic Physics, Hardenbergstr. 36, 10623 Berlin, Germany}

\author{Norman Vincenz Ewald}
\affiliation{German Aerospace Center (DLR), Institute of Optical Sensor Systems, Rutherfordstr. 2, 12489 Berlin, Germany}

\author{Mustafa Gündo\u{g}an}
\affiliation{German Aerospace Center (DLR), Institute of Optical Sensor Systems, Rutherfordstr. 2, 12489 Berlin, Germany}
\affiliation{Institut für Physik, Humboldt-Universit\"at zu Berlin, Newtonstr. 15, 12489 Berlin, Germany}

\author{Janik Wolters}
\affiliation{German Aerospace Center (DLR), Institute of Optical Sensor Systems, Rutherfordstr. 2, 12489 Berlin, Germany}
\affiliation{TU Berlin, Institute for Optics and Atomic Physics, Hardenbergstr. 36, 10623 Berlin, Germany}


\date{\today}

\begin{abstract}
Quantum memories promise to enable global quantum repeater networks. For field applications, alkali metal vapors constitute an exceptional storage platform, as neither cryogenics, nor strong magnetic fields are required. We demonstrate a technologically simple, in principle satellite-suited quantum memory based on electromagnetically induced transparency on the cesium D1 line, and focus on the trade-off between end-to-end efficiency and signal-to-noise ratio, both being key parameters in applications. For coherent pulses containing one photon on average, we achieve storage and retrieval with end-to-end efficiencies of $\eta_{\text{e2e}}=13(2)\%$, which correspond to internal memory efficiencies of $\eta_{\text{mem}}=33(1)\%$.  Simultaneously, we achieve a  noise level corresponding to $\mu_1=0.07(2)$ signal photons. This noise is dominated by spontaneous Raman scattering, with contributions from fluorescence. Four wave mixing noise is negligible, allowing for further minimization of the total noise level.
\end{abstract}


\maketitle



Quantum key distribution protocols \cite{Bennett.2014, Ekert.1991} harness superposition and entanglement to ensure secure information transfer between two parties \cite{Gisin.2002}. However, the longstanding bottleneck of long-distance quantum communication are photonic losses. The quantum repeater (QR) was proposed in Ref.$\,$\cite{Briegel.1998} as a general solution to this problem by dividing a long-distance link into segments, and making use of entanglement swapping between the nodes to generate entanglement over the long-distance link. An essential requirement for QRs for ground- and satellite-based quantum networks \cite{Gundogan.2021, Wallnofer.2022, Liorni.2021} are quantum memories (QM) \cite{Lvovsky.2009, Heshami.2016}. These act as interfaces between flying and stationary qubits and allow for storage of quantum information, ideally in an efficient and noise-free manner. QMs are being developed in many different platforms, ranging from ultracold atoms \cite{Bao.2012, YangSJ.2016, Pu.2017} to solid-state systems \cite{Afzelius.2009, Gundogan.2015, YangTS.2018, Ma.2021, Ortu.2022} and warm atomic vapors \cite{Saunders.2016, Wolters.2017, Katz.2018, Guo.2019, Thomas.2019}. The latter constitute technologically simple and scalable systems, more suitable for use on satellites, as neither laser cooling, strong magnetic fields nor ultra-high vacuum are needed for their implementation. Due to a high optical depth, high efficiencies are possible. However, these types of memories are often considered to be prone to noise, especially four-wave-mixing (FWM) \cite{Michelberger.2015}. \\
%
\begin{figure*}[t]
\includegraphics[trim={0 10.5cm 0 0}, clip, scale=0.55]{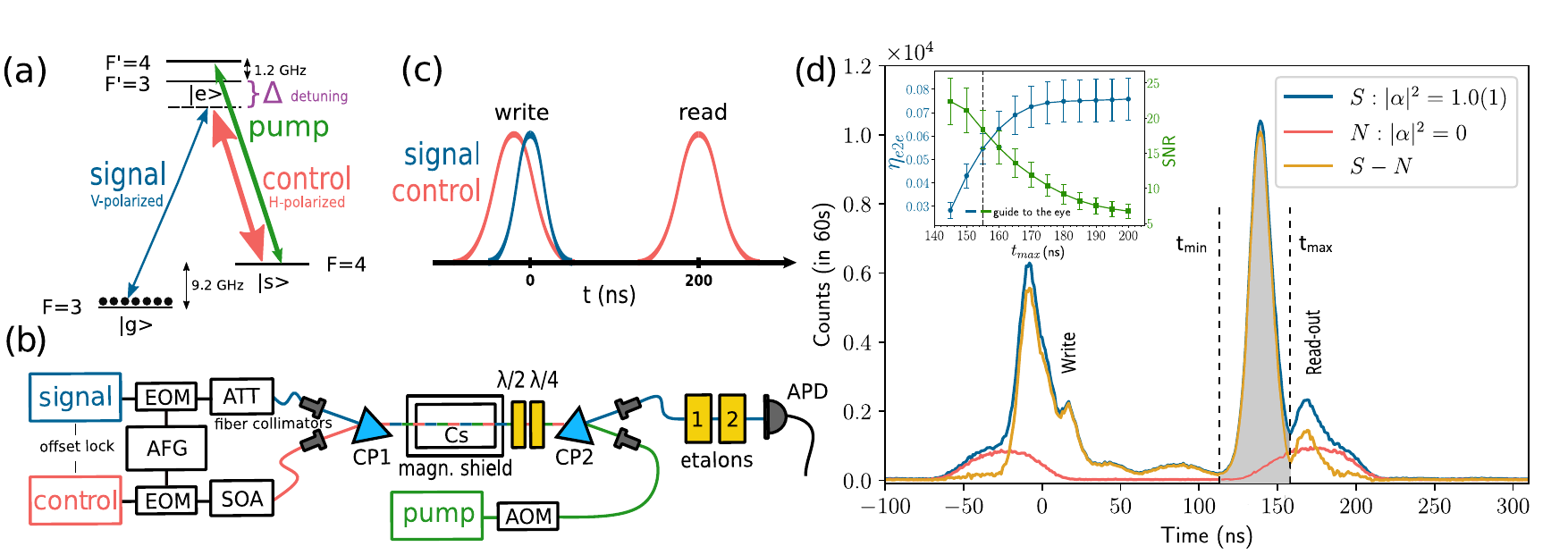}
\caption{\label{setup&storage} Overview of the experiment. (a) Three-level $\Lambda$-system with signal (control) laser frequency red-detuned by $\Delta$ from the respective hyperfine transitions of the Cs D1 line. A pump laser initially prepares the $F=3$ ground state ($\ket{g}$). Both ground states $\ket{g}$ and $\ket{s}$ are naturally long-lived. (b) Schematic memory-experiment setup. EOM: electro-optic modulator; AFG: arbitrary function generator; ATT: attenuator; SOA: semiconductor optical amplifier; CP1, CP2: calcite prisms; Cs: vapor cell inside magnetic shielding; $\lambda/2$, $\lambda/4$: wave plates;
APD: single-photon counting avalanche photodiode; AOM: acousto-optic modulator. (c) Signal and control pulse sequence (not to scale). The signal pulse enters the memory shortly after the first control pulse and is written into the memory by it, while the second control pulse reads out the information 80-200$\,$ns later. \\ (d) Exemplary arrival time histogram of detected photons in a memory experiment integrated for $60\,$s for an incoming coherent state with a Gaussian envelope containing $\abs{\alpha}^2=1.0(1)$ photons on average (blue), for blocked input signal ($\abs{\alpha}^2=0$) (red), as well as the resulting noise-corrected signal (yellow). The storage time is $t_\text{storage}=140(1)\,$ns as measured from the maximum of the input signal to the global maximum of the retrieval pulse. The shaded area shows the detection window of width $t_{\text{max}}-t_{\text{min}}$ used for further analysis of the end-to-end efficiency $\eta_{\text{e2e}}$ of the memory setup and the signal-to-noise ratio SNR. We use $t_{\text{max}}= 155\,$ns (black dashed line), being a good compromise between $\eta_{\text{e2e}}$ and SNR (Inset). } 
\end{figure*}
Building secure global quantum communication networks \cite{Kimble.2008} will make it necessary to pair such memories with bright and efficient single-photon sources such as semiconductor quantum dots (QDs) \cite{Tang.2015, Somaschi.2016, Ding.2016, Beguin.2018, Kroh.2019}, spontaneous parametric down-conversion (SPDC) sources \cite{Schunk.2015, Seri.2017, Mottola.2020, Buser.2022}, or sources based on room temperature atomic ensembles \cite{Dideriksen.2021, Davidson.2021}. Essential requirements on compatible quantum memories for this endeavor are high end-to-end efficiencies $\eta_{\text{e2e}}$, high signal-to-noise ratios (SNR), and sufficiently long storage times in the millisecond range. In order to benchmark the memory performance, we use the parameter $\mu_1$, defined as the mean photon number in the input that results in a SNR of 1 in the output \cite{Jobez.2015}. In Refs.$\,$\cite{Gundogan.2021, Wallnofer.2022}, key rates for specific QR configurations are simulated with realistic memory parameters, such as memory efficiencies of $\eta_{\text{mem}}=70-80\%$ and storage times of some milliseconds. Previously, an internal memory efficiency of $\eta_{\text{mem}}=80\%$, without considering the filtering system $\,$\cite{Guo.2019}, and a low noise figure of $\mu_1=0.20(2)$ \cite{Thomas.2019} have been reported. A $1/e$ storage time of $t_\text{store}=430\,$ms for attenuated coherent pulses was also achieved $\,$\cite{Katz.2018}. Nevertheless, in these state-of-the-art implementations only one parameter $(\eta_{\text{e2e}},\, \mu_1,\, t_\text{store})$ was optimized at a time. Especially regarding applications in quantum communication, it is imperative to find operating conditions which maximize all relevant parameters simultaneously. \\
Our work focuses on the simultaneous optimization of $\eta_{\text{e2e}}$ and $\mu_1$. The used EIT $\Lambda$-configuration with copropagating signal and control is in principle suitable for long storage times above $1\,$ms, which will be pursued in follow-up experiments. We achieve $\eta_{\text{e2e}}=13(2)\%$ and $\mu_1=0.07(2)$ simultaneously. The internal memory efficiency reaches $\eta_{\text{mem}}=33(1)\%$. The origin of the read-out noise is found to be a combination of spontaneous Raman scattering (SRS) and fluorescence noise, where the former dominates at larger control pulse energies. Apart from practical applications in long-distance QKD, these results are of general interest for the community involved in the exploitation of laser-induced atomic coherence. \\
%
The presented hyperfine ground-state memory scheme is based on EIT \cite{Fleischhauer.2005, Phillips.2001} on the Cs D1 transition, following the scheme shown in Fig.$\,$\ref{setup&storage}(a), and is, to our knowledge, the first realization of a warm single-photon-level memory in cesium on this transition, although there exist realizations using cold atoms, e.g. Ref.~\cite{Tseng.2022}. Cesium offers a higher possible bandwidth, due to a larger ground-state hyperfine splitting, as compared to e.g.\ rubidium. Comparatively, cesium also shows a larger excited state hyperfine splitting, which reduces the influence of the additional, unwanted excited state. Further advantages of cesium include its larger mass and longer D1 wavelength, which result in slower thermal movement and therefore less Doppler broadening, and in less absorption in optical fibers compared to the Rb D1 line, respectively. As depicted in the experimental setup in Fig.$\,$\ref{setup&storage}(b), the orthogonal and linearly polarized signal ($S$) and control ($C$) lasers (ECDLs from Sacher Lasertechnik) are red-detuned by $\Delta$ from the $F\,=3\rightarrow\,F'=3$ and $F\,=4\rightarrow\,F'=3$ transitions respectively, and are offset-locked to a frequency difference of $\sim 9.2\,$ GHz. A small deviation from the exact hyperfine splitting is caused by the AC-Stark shift induced by the control laser. Gaussian signal and control pulses of varying full width at half maximum (FWHM) are generated by fiber-based electro-optic modulators controlled by arbitrary function generators. The control pulses are amplified by a semiconductor optical amplifier (SOA). The SOAs amplified spontaneous emission is suppressed by a combination of a narrow-band dielectric filter ($1\,$nm FWHM) and a Fabry-Pérot etalon (free spectral range FSR $=205.5\,$GHz, finesse $\mathcal{F}=47$). This results in a control laser peak power of $P_{C,\text{max}}=12.9\,$mW. For single-photon-level measurements the signal pulses are attenuated with optical filters. To calibrate the photon number, $10\%$ of the signal pulse ($N_{\text{mon}}$) is monitored on an avalanche photodiode (APD). The signal and control beams are overlapped using a calcite prism and sent through the Cs memory cell. Hereby, we put special care into making signal and control beams co-propagate through the cylindrical vapor cell ($7.5\,$cm length, $2\,$cm diameter). The cell contains ${}^{133}$Cs and $5\,$Torr N$_2$ buffer gas, is enclosed inside a $\mu$-metal magnetic shielding to nominally reduce the oscillating magnetic field in its interior by a factor of $1000$, and is kept at $T_\text{cell}=60\,$°C. We estimate the wavelength of the spin wave due to the residual angle to be $\lambda_\text{sw}=2\pi/|\vec{k}_S-\vec{k}_C|\approx 5\,$mm. The control and signal beam diameters are $109(5)\,\mu$m and $93(5)\,\mu$m FWHM at their focus at the center of the memory cell, and their Rayleigh length (as for all beams) extends until the cell's end. With a dipole moment of $d=2.7 \cdot 10^{-29}\,$C$\cdot$m \cite{Steck.2019} for the Cs D1 transition, this yields a peak control Rabi frequency of $\Omega_C \approx 2\pi\cdot 540\,$MHz. A pump laser on the $F\,=4\rightarrow\,F'=4$ transition, turned on and off using an acousto-optic modulator (AOM), is used for state preparation in the $F\,=3$ ground state ($\ket{g}$). Its beam diameter is $540(5)\,\mu$m FWHM at its focus, with a peak power of $P_{\text{pump}}=12.3\,$mW. The measured pumping efficiency at this configuration is approximately $n_{F=3}/n_{F=4}=80\%$.\\
To filter the signal from the control pulses we employ a combination of polarization- and spectral filtering. The former yields a suppression by 7 orders of magnitude for the control light. The following spectral filtering is accomplished by two Fabry-Pérot etalons (FSR$\,=25.6\,$GHz, $\mathcal{F}=47$), each of which further attenuates the control laser by $3$ orders of magnitude within the signal beam path. In total, this amounts to a control laser suppression by $13$ orders of magnitude. Meanwhile, the signal light is only attenuated by a factor of $2.5$. At the output, the signal photons are detected by a single-photon-counting APD (Laser Components COUNT-50C-FC) and the resulting signal is analyzed using time-correlated single photon counting. \\
%
\begin{figure*}[t]
\includegraphics[trim={0 13cm 0 0}, clip, scale=0.6]{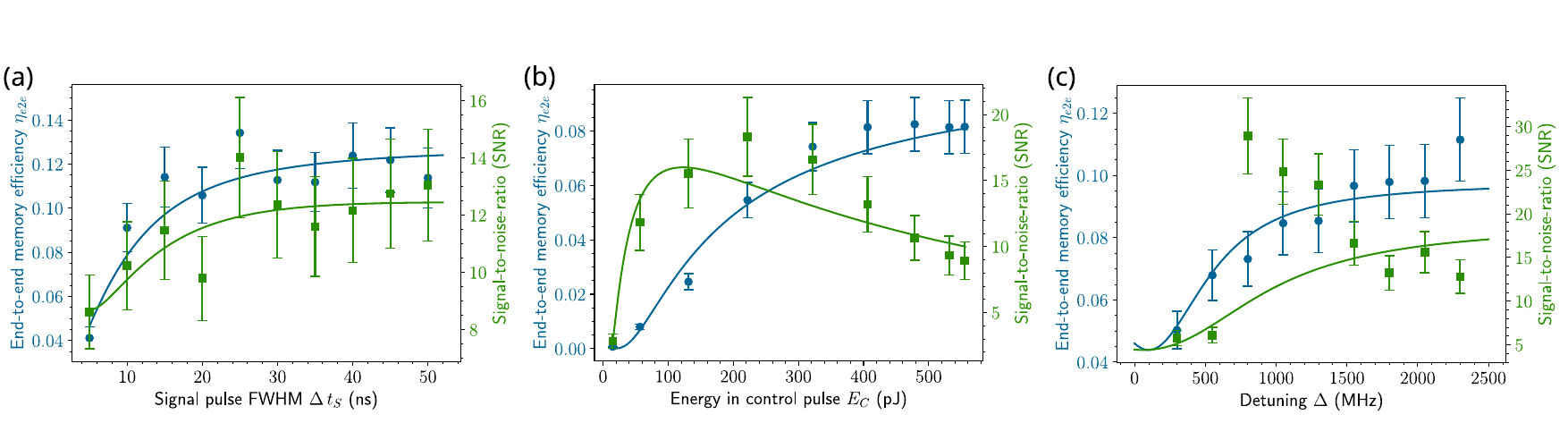}
\caption{ \label{eff_SNR} 
End-to-end efficiency ($\eta_{\text{e2e}}$) and signal-to-noise ratio (SNR). $\eta_{\text{e2e}}$ (blue circles) and SNR (green squares) of the memory as a function of signal pulse width $\Delta t_S$ (a), control pulse energy $E_C$ (b), and laser detuning $\Delta$ (c). The values selected for the measurements are $\Delta t_S=25\,$ns, $E_C = 560(50)\,$pJ, and $\Delta=2300(100)\,$MHz red detuning from the atomic resonance, with only the respective parameter being varied.}
\end{figure*}
A single run of the storage-retrieval experiment begins by switching on the pump laser for $10\,\mu$s to prepare the state $\ket{g}$. After pumping, the control and signal pulses enter the memory. Hereby, the signal pulse is delayed for several ns with respect to the control pulse, as depicted in Fig.$\,$\ref{setup&storage}(c). To retrieve the signal photon from the memory, a second identical control pulse is sent into the cell after $t=200\,$ns. The storage and retrieval experiment is repeated at a rate $f_{\text{rep}}=1/(11\mu$s) for an integration time $ t_{\text{int}}$ of $60\,$s. \\
%
Figure$\,$\ref{setup&storage}(d) shows a typical measured photon arrival time histogram as used for all further analysis. Shown are measured data for storage and retrieval of a coherent state containing  $\abs{\alpha}^2=(N_{\text{mon}}\,\sigma)/(f_\text{rep}\,\eta_\text{APD})=1.0(1)$ photons on average, a noise measurement resulting from storage of a vacuum state with $\abs{\alpha}^2=0$, arising from a blocked input, and the noise-corrected signal. The uncertainty of $\abs{\alpha}^2$ arises from the error in the signal/monitor splitting ratio $\sigma$, measured with a power-meter (Thorlabs PM160), the uncertainty of the APD efficiency $\eta_{\text{APD}}=0.33(5)$, and the statistical error in the monitored photon number $N_{\text{mon}}$. The peak at $t=0\,$ns corresponds to the part of the signal pulse that is transmitted through the memory (leakage). The peak at $t=140\,$ns corresponds to the signal read out from the memory. As shown in Figure$\,$\ref{setup&storage}(d), the temporal profile of the retrieved signal peak is distorted by the memory and shows two distinct maxima. We define the storage time $t_\text{storage}$ as the difference between the first (and highest) of these maxima-, and the peak of the incoming signal at $t=0\,$ns. The used storage times were chosen to be short in order to focus on the $\eta_{\text{e2e}}$ and SNR, but can be extended into the $\mu$s range with the present setup, and into the millisecond range by increasing the beam diameters and changing the $\Lambda$-system configuration by exploiting the magnetic Zeeman sublevels of the hyperfine ground states \cite{Katz.2018}. \\
Given the temporal envelope of the retrieved signal, the $\eta_{\text{e2e}}$ of the memory setup, including the filtering system, and the SNR at single photon level are given as 

\begin{align}
\eta_{\text{e2e}} &= \frac{N_{\text{signal}}-N_{\text{noise}}}{\abs{\alpha}^2\, \eta_{\text{APD}}\, f_{\text{rep}}\, t_{\text{int}}}, \label{eq_eta}\\
\text{SNR}_{\alpha} &= \frac{N_{\text{signal},\alpha}-N_{\text{noise}}}{N_{\text{noise}}} \label{eq_SNR},
\end{align}

\noindent
where $N_{\text{signal},\alpha}$($N_{\text{noise}}$) correspond to the number of signal(noise) counts. Due to the long integration time, the statistical error is small and the uncertainty in $\eta_{\text{e2e}}$ is dominated by the systematic error in the power measurement and the calibration of the single photon level. For the SNR, as it is measured at single-photon level, the error bars are proportional to the uncertainty of $\abs{\alpha}^2$. The $\eta_{\text{e2e}}$ increases with the length of the retrieval time integration window $t_\text{max}$, while the SNR decreases, as can be taken from the inset in Fig.$\,$\ref{setup&storage}(d). Hence, a compromising value of $t_{\text{max}}=155\,$ns was chosen. The lower integration limit is always taken at the minimum before each retrieval peak.  
%
\begin{figure*}[t!]
\includegraphics[trim={0 12cm 0 0}, clip, scale=0.605]{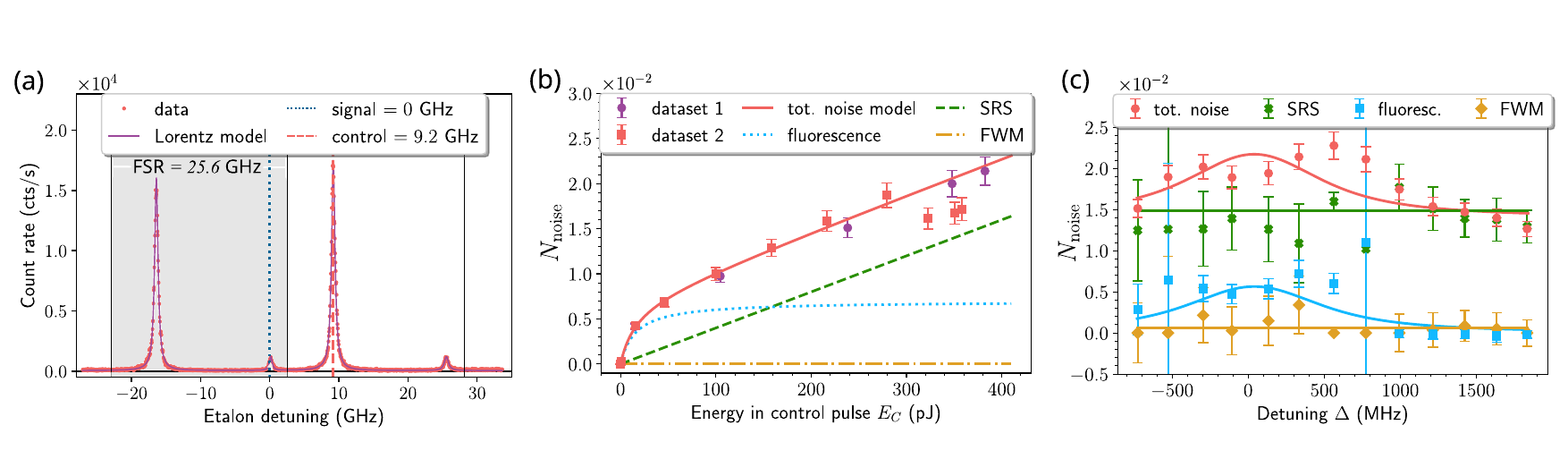}
\caption{ \label{Noise} Noise analysis. (a) Noise detected on the APD, while scanning one of the etalons used for spectral filtering over two free spectral ranges (FSR) (shaded area corresponds to one FSR), while keeping the other etalon on resonance with the signal frequency. Large peaks at $-16.4\,$GHz and $9.2\,$GHz detuning correspond to leaked control-laser light, and small peaks close to the signal frequency (and at $25.6\,$GHz detuning) originate from atomic noise (fluorescence, SRS and FWM). Vertical dashed lines indicate the signal (blue) and control (red) frequencies and vertical black solid lines indicate the limits of one FSR. \\(b) Exemplary measurement of total noise counts (the two different datasets were taken on different days) per retrieval attempt as a function of the energy in the control pulse $E_C$ for $\Delta = 334\,$MHz (red detuning), and corresponding fits of the different noise components: fluorescence, spontaneous Raman scattering (SRS), and four-wave mixing (FWM). (c) Systematic study of the measured total noise counts and the corresponding fitted noise components at retrieval as a function of the control laser detuning $\Delta$. Negative values of $\Delta$ correspond to blue detuning from resonance. The noise is mainly dominated by SRS and florescence, while FWM is negligible in this setting.}
\end{figure*}
%
Figure \ref{eff_SNR} shows the dependence of $\eta_{\text{e2e}}$ and SNR on signal pulse width $\Delta t_S$ (a), energy in the control pulses $E_C$ (b), and detuning $\Delta$ (c). $\Delta t_S\,$  is related to the bandwidth 
 $\Delta \omega_S$ (FWHM) of the pulses that are stored in the memory as $\Delta t_S\,\Delta \omega_S = 2\,ln(2)/\pi$, for transform-limited pulses. For the measurements we took $\Delta t_S=25\,$ns and $\Delta=2300(100)\,$MHz as baseline and varied only the respective parameter. $E_C$ was hereby adjusted in a range $280(50)\,\text{pJ}\leq E_C \leq 560(50)\,$pJ. In all three cases, $\eta_{\text{e2e}}$ shows a saturation behavior, while the SNR also saturates in Fig. \ref{eff_SNR}(a), but reaches a maximum in Figs. \ref{eff_SNR}(b) and (c), and then decreases again. Therefore, there exists a trade-off between reaching the maximally possible efficiency, and adding more noise to the retrieval signal. We note that the large scatter in the SNR visible in Fig.~\ref{eff_SNR}(c) was due to imperfect measurement in the early stage of the experiment. With more accurate measurements performed later on, monitoring the laser frequency using a wavemeter (Wavelength Meter WS/7-60), such a large scatter was not visible anymore. \\
To describe $\eta_{\text{e2e}}$, the following models were used: The efficiency as a function of signal pulse width (Fig.~\ref{eff_SNR}(a)) is $\eta_{\text{e2e}}(\Delta t_S)={\eta}_{0,t_S}\,/\sqrt{1+ ( \frac{4\,ln(2)}{ \Delta t_S\, \Delta \omega_{\text{mem}}} )^2 }$, where $\eta_{0,t_S}$ is the maximal achievable efficiency and $\Delta\omega_{\text{mem}}$ is the bandwidth of the memory. This model arises from the Maxwell-Bloch equations  and describes how efficiently a pulse of given width $\Delta t_S\,$ is converted into an atomic spin wave, when limited by the available control Rabi frequency $\Omega_C$, which determines the spectral width of the EIT window, and therefore the memory bandwidth \cite{Hsiao.2018}. It is in good agreement with the measured data and yields ${\eta}_{0,t_S}=0.128(7)$ and $\Delta\omega_{\text{mem}}=220(30)\,$MHz.
In Fig.~\ref{eff_SNR}(b) we model $\eta_{\text{e2e}}(E_C)=\eta_{0,C}\,e^{-a/E_C}$, motivated by the dependence on the control pulse energy of the former model. For the fit parameters we find $\eta_{0,C}=0.107(7)$ and $a=1.56(9)\cdot 10^{2}\,\text{pJ}$. Due to non-ideal etalon adjustments, $\eta_{0,C}$ is slightly reduced compared to $\eta_{\text{e2e}}$. The dependence on the detuning $\Delta$ shown in Fig.$\,$\ref{eff_SNR}(c) is modeled by $\eta_{\text{e2e}}(\Delta)=\eta_{0,\Delta}\,e^{-\alpha(\Delta)}$, where $\alpha(\Delta)$ is a Lorentzian, reflecting that the memory efficiency is limited by collision-broadened absorption from the atomic ensemble. These models are not to be taken as a result from a microscopic theory, but more as a hint to the underlying physical mechanisms. A full theoretical model is outlined in Refs.$\,$\cite{Gorshkov.2007, Gorshkov.2007a, Gorshkov.2007b, Gorshkov.2007c, Gorshkov.2008d, Rakher.2013}. The models for the SNR are determined as the ratio between the efficiency- and the noise models, the latter of which are described in the next section. \\
%
%
To understand the limiting noise sources, a thorough experimental noise analysis is performed. When the temporal, spatial, and spectral leakage of the control light is sufficiently suppressed, the three most relevant noise sources in warm vapor quantum memories are fluorescence, SRS, and FWM \cite{Heshami.2016}. 
Figure \ref{Noise}(a) shows the frequency dependence of the measured noise count rate, obtained by scanning the second etalon of the filtering system over two FSR, each of $25.6\,$GHz, by temperature tuning, while keeping the first etalon on resonance with the signal frequency. This measurement is taken with blocked signal laser. Two large peaks at $-16.4\,$GHz and $9.2\,$GHz detuning appear, corresponding to leaked control laser light. The transmitted control laser power is negligible when the etalon is in resonance with the signal, i.e.~during storage experiments. Additionally, two smaller peaks close to the signal frequency appear approximately at $0$ and $25.6\,$GHz detuning. This indicates that the only relevant noise sources during memory operation are of atomic origin. In order to attain a better understanding of these, we measured the number of noise photons $N_\text{noise}$ counted on the APD by varying $\Delta$, while monitoring the control laser frequency with the wavemeter. Here, $\Delta=0(60)\,$MHz corresponds to the control transition. At each point, we determined the noise count dependence on $E_C$, yielding a two dimensional analysis.\\
Figure $\,$\ref{Noise}(b) depicts the noise counts as a function of the control pulse energy $E_C$ at $\Delta = 334\,$MHz (red detuning). The data can be modeled by a second-order polynomial plus a saturating component, $N_\text{noise}=b E_C^2+c E_C + d E_C / (e + E_C) $, where the quadratic and linear contributions account for FWM and SRS respectively, and the saturating part corresponds to fluorescence noise. For the shown detuning value we find $b=0(10^{-8})\,(\text{pJ})^{-2}$, $c=4(2)\cdot 10^{-5}\,(\text{pJ})^{-1}$, $d=7(2)\cdot 10^{-3}\,(\text{pJ})^{-1}$, and $e=16(16)\,\text{pJ}$, indicating that both fluorescence and SRS noise have a significant contribution, while FWM vanishes. \\
%
Figure \ref{Noise}(c) depicts the behavior of the noise for varying detuning $\Delta$ at maximal $E_C$. The error bars of the total noise are of statistical origin. From fitting the different noise components as a function of the control field energy at each detuning value (as in Figure \ref{Noise}(b)), the corresponding components fluorescence, SRS, and FWM are determined at each point. Here, the error bars arise from the fit error of the $E_C$ dependence. The error bars at $\Delta=-526~\text{and}~774\,$MHz are comparatively large, since the fit hardly converged and FWM was manually excluded for these data points. From the analysis we assume that FWM is negligible in the experiment, SRS yields mainly a constant contribution to the noise, and the detuning dependence of the total noise arises mainly from fluorescence. The fluorescence was modeled as a Voigt distribution with a Gaussian component of $380\,$MHz width at FWHM due to Doppler broadening, and a Lorentzian component of $920\,$MHz at FWHM, arising from power broadening by the control light and pressure broadening due to collisions with the buffer gas, with the only free parameter being the height. This simple model conveys the general trend and is valid for the here investigated small detunings on the order of the broadened atomic transition. At large detunings, a decay of SRS is expected, but even for the largest here measured detuning values it is not pronounced enough to cause significant deviations from the model. 
The negligible FWM contribution is particularly interesting, as this type of noise is commonly understood to be the main drawback of vapor cell quantum memories and efforts are made for its active mitigation \cite{Saunders.2016, Thomas.2019}. As we operate with low bandwidth pulses, moderate control powers, and at the D1 line, we manage to practically suppress this type of noise. At detunings close to zero, a local minimum appears in the noise data. We presume this minimum to be caused by re-absorption of the light due to residual spatial miss-match of the beams along the whole length of the vapor cell. A more detailed model considering this re-absorption, the decay of SRS at large detunings, and the slight fluctuations of the noise data has too many free parameters to fit the number of existing data points. More investigation would be required to fully understand this fact, and will be pursued in further experiments and theoretical calculations. The resulting total noise model $N_\text{noise}(\Delta)=N_{\text{SRS}} + N_{\text{fl}}V(\Delta)+N_{\text{FWM}}$ is then a sum of the constant SRS contribution and the Voigt distribution resulting from fluorescence. For completeness, we also add the FWM contribution to the model. The resulting parameters are $N_{\text{SRS}}=14(1)\cdot 10^{-3}$, $N_{\text{fl}}=7(2)\cdot 10^{-3}$, and $N_{\text{FWM}}=0(1)\cdot 10^{-3}$. It is thus essential to reduce the former two for future optimization.\\
Combining the models for $\eta_{\text{e2e}}$ with the noise models yields the SNR models shown in Fig.$\,$\ref{eff_SNR}(a)-(c). These are in reasonable agreement with the measured data, and confirm the applicability of the former models. \\
In summary, we have realized, to the best of our knowledge, the first single-photon-level storage in a warm hyperfine EIT cesium memory on the Cs D1 line with an end-to-end efficiency of $\eta_{\text{e2e}}=(13\pm 2)\%$, which translates to an internal memory efficiency of $\eta_{\text{mem}}=(33\pm 1)\%$. These values approach those needed for practical implementations on QRs \cite{Gundogan.2021, Wallnofer.2022}. The simultaneously obtained SNR is $14\pm2$, which can be converted to a noise level corresponding to $\mu_1=0.07(2)$ signal photons. As the limiting noise sources, SRS and fluorescence are identified. The present memory is not yet fully optimized for long storage times, but the geometry with parallel signal and control beams in principle allows for storage times not limited by motional dephasing, at least in the ms regime. Nonetheless, our device can already find a use in quantum computation and sensing applications where deterministic generation of multiphoton states is required \cite{Nunn.2013}. The used control powers in this experiment were comparatively low. However, much higher powers on the order of hundreds of mW will be needed in future experiments, where larger beam sizes are planned, in order to further extend the storage times. For minimizing SRS noise, the pumping efficiency should be increased using higher pump laser powers and larger beams. Further noise reduction is planned by exploiting the Zeeman-substructure of the cesium hyperfine states and preparing the atoms in the $m_F=+4$ sublevel with $\sigma^+$ polarized light, thus suppressing residual FWM by selection rules, and reducing possible SRS noise contributions \cite{Guan.2007, Xu.2013, Katz.2018}. In this new configuration, storage times of a few milliseconds are within reach. They will thus be extended to reach the benchmark $t_\text{store}=1\,$ms for practical applications in QRs, by increasing the beam diameter up to the cell diameter, hence inhibiting atomic diffusion out of the optical interaction volume.  By use of optimal control pulses as in Ref.~\cite{Guo.2019}, we will simultaneously boost the memory efficiency even further. With this, end-to-end efficiencies above $70\%$ and simultaneously even higher signal-to-noise ratios, as well as ms-long storage times are within reach, laying the foundations for memory-assisted, satellite-suited QRs for long-distance quantum communication. \\

\begin{acknowledgments}
The authors kindly thank Gianni Buser for interesting and fruitful discussions. We acknowledge the support by the German Space Agency (DLR) with funds provided by the Federal Ministry of Economics and Technology (BMWi) under grant number 50RP2090 (QuMSeC) and by the Federal Ministry of Education and Research (BMBF) under grant number 16KISQ040K (Q-ToRX). M.G.$\,$ acknowledges funding from the European Union’s Horizon 2020 research and innovation program under the Marie Sklodowska-Curie grant agreement No 894590. E.R. acknowledges funding through the Helmholtz Einstein International Berlin Research School in Data Science (HEIBRiDS)
\end{acknowledgments}

\bibliography{Paper1_Luisa_R}

\begin{thebibliography}{50}%
\makeatletter
\providecommand \@ifxundefined [1]{%
 \@ifx{#1\undefined}
}%
\providecommand \@ifnum [1]{%
 \ifnum #1\expandafter \@firstoftwo
 \else \expandafter \@secondoftwo
 \fi
}%
\providecommand \@ifx [1]{%
 \ifx #1\expandafter \@firstoftwo
 \else \expandafter \@secondoftwo
 \fi
}%
\providecommand \natexlab [1]{#1}%
\providecommand \enquote  [1]{``#1''}%
\providecommand \bibnamefont  [1]{#1}%
\providecommand \bibfnamefont [1]{#1}%
\providecommand \citenamefont [1]{#1}%
\providecommand \href@noop [0]{\@secondoftwo}%
\providecommand \href [0]{\begingroup \@sanitize@url \@href}%
\providecommand \@href[1]{\@@startlink{#1}\@@href}%
\providecommand \@@href[1]{\endgroup#1\@@endlink}%
\providecommand \@sanitize@url [0]{\catcode `\\12\catcode `\$12\catcode
  `\&12\catcode `\#12\catcode `\^12\catcode `\_12\catcode `\%12\relax}%
\providecommand \@@startlink[1]{}%
\providecommand \@@endlink[0]{}%
\providecommand \url  [0]{\begingroup\@sanitize@url \@url }%
\providecommand \@url [1]{\endgroup\@href {#1}{\urlprefix }}%
\providecommand \urlprefix  [0]{URL }%
\providecommand \Eprint [0]{\href }%
\providecommand \doibase [0]{https://doi.org/}%
\providecommand \selectlanguage [0]{\@gobble}%
\providecommand \bibinfo  [0]{\@secondoftwo}%
\providecommand \bibfield  [0]{\@secondoftwo}%
\providecommand \translation [1]{[#1]}%
\providecommand \BibitemOpen [0]{}%
\providecommand \bibitemStop [0]{}%
\providecommand \bibitemNoStop [0]{.\EOS\space}%
\providecommand \EOS [0]{\spacefactor3000\relax}%
\providecommand \BibitemShut  [1]{\csname bibitem#1\endcsname}%
\let\auto@bib@innerbib\@empty
\bibitem [{\citenamefont {Bennett}\ and\ \citenamefont
  {Brassard}(2014)}]{Bennett.2014}%
  \BibitemOpen
  \bibfield  {author} {\bibinfo {author} {\bibfnamefont {C.~H.}\ \bibnamefont
  {Bennett}}\ and\ \bibinfo {author} {\bibfnamefont {G.}~\bibnamefont
  {Brassard}},\ }\href {https://doi.org/\url{10.1016/j.tcs.2014.05.025}}
  {\bibfield  {journal} {\bibinfo  {journal} {Theor. Comput. Sci.}\ }\textbf
  {\bibinfo {volume} {560}},\ \bibinfo {pages} {7} (\bibinfo {year}
  {2014})}\BibitemShut {NoStop}%
\bibitem [{\citenamefont {Ekert}(1991)}]{Ekert.1991}%
  \BibitemOpen
  \bibfield  {author} {\bibinfo {author} {\bibfnamefont {A.~K.}\ \bibnamefont
  {Ekert}},\ }\href@noop {} {\bibfield  {journal} {\bibinfo  {journal} {Phys.
  Rev. Lett.}\ }\textbf {\bibinfo {volume} {67}},\ \bibinfo {pages} {661}
  (\bibinfo {year} {1991})}\BibitemShut {NoStop}%
\bibitem [{\citenamefont {Gisin}\ \emph {et~al.}(2002)\citenamefont {Gisin},
  \citenamefont {Ribordy}, \citenamefont {Tittel},\ and\ \citenamefont
  {Zbinden}}]{Gisin.2002}%
  \BibitemOpen
  \bibfield  {author} {\bibinfo {author} {\bibfnamefont {N.}~\bibnamefont
  {Gisin}}, \bibinfo {author} {\bibfnamefont {G.}~\bibnamefont {Ribordy}},
  \bibinfo {author} {\bibfnamefont {W.}~\bibnamefont {Tittel}},\ and\ \bibinfo
  {author} {\bibfnamefont {H.}~\bibnamefont {Zbinden}},\ }\href@noop {}
  {\bibfield  {journal} {\bibinfo  {journal} {Rev. Mod. Phys.}\ }\textbf
  {\bibinfo {volume} {74}},\ \bibinfo {pages} {145} (\bibinfo {year}
  {2002})}\BibitemShut {NoStop}%
\bibitem [{\citenamefont {Briegel}\ \emph {et~al.}(1998)\citenamefont
  {Briegel}, \citenamefont {D{\"u}r}, \citenamefont {Cirac},\ and\
  \citenamefont {Zoller}}]{Briegel.1998}%
  \BibitemOpen
  \bibfield  {author} {\bibinfo {author} {\bibfnamefont {H.-J.}\ \bibnamefont
  {Briegel}}, \bibinfo {author} {\bibfnamefont {W.}~\bibnamefont {D{\"u}r}},
  \bibinfo {author} {\bibfnamefont {J.~I.}\ \bibnamefont {Cirac}},\ and\
  \bibinfo {author} {\bibfnamefont {P.}~\bibnamefont {Zoller}},\ }\href
  {https://doi.org/\url{10.1103/PhysRevLett.81.5932}} {\bibfield  {journal}
  {\bibinfo  {journal} {Phys. Rev. Lett.}\ }\textbf {\bibinfo {volume} {81}},\
  \bibinfo {pages} {5932} (\bibinfo {year} {1998})}\BibitemShut {NoStop}%
\bibitem [{\citenamefont {G{\"u}ndo{\u{g}}an}\ \emph
  {et~al.}(2021)\citenamefont {G{\"u}ndo{\u{g}}an}, \citenamefont {Sidhu},
  \citenamefont {Henderson}, \citenamefont {Mazzarella}, \citenamefont
  {Wolters}, \citenamefont {Oi},\ and\ \citenamefont
  {Krutzik}}]{Gundogan.2021}%
  \BibitemOpen
  \bibfield  {author} {\bibinfo {author} {\bibfnamefont {M.}~\bibnamefont
  {G{\"u}ndo{\u{g}}an}}, \bibinfo {author} {\bibfnamefont {J.~S.}\ \bibnamefont
  {Sidhu}}, \bibinfo {author} {\bibfnamefont {V.}~\bibnamefont {Henderson}},
  \bibinfo {author} {\bibfnamefont {L.}~\bibnamefont {Mazzarella}}, \bibinfo
  {author} {\bibfnamefont {J.}~\bibnamefont {Wolters}}, \bibinfo {author}
  {\bibfnamefont {D.~K.~L.}\ \bibnamefont {Oi}},\ and\ \bibinfo {author}
  {\bibfnamefont {M.}~\bibnamefont {Krutzik}},\ }\href
  {https://doi.org/\url{10.1038/s41534-021-00460-9}} {\bibfield  {journal}
  {\bibinfo  {journal} {npj Quantum Information}\ }\textbf {\bibinfo {volume}
  {7}},\ \bibinfo {pages} {128} (\bibinfo {year} {2021})}\BibitemShut {NoStop}%
\bibitem [{\citenamefont {Walln{\"o}fer}\ \emph {et~al.}(2022)\citenamefont
  {Walln{\"o}fer}, \citenamefont {Hahn}, \citenamefont {G{\"u}ndo{\u{g}}an},
  \citenamefont {Sidhu}, \citenamefont {Wiesner}, \citenamefont {Walk},
  \citenamefont {Eisert},\ and\ \citenamefont {Wolters}}]{Wallnofer.2022}%
  \BibitemOpen
  \bibfield  {author} {\bibinfo {author} {\bibfnamefont {J.}~\bibnamefont
  {Walln{\"o}fer}}, \bibinfo {author} {\bibfnamefont {F.}~\bibnamefont {Hahn}},
  \bibinfo {author} {\bibfnamefont {M.}~\bibnamefont {G{\"u}ndo{\u{g}}an}},
  \bibinfo {author} {\bibfnamefont {J.~S.}\ \bibnamefont {Sidhu}}, \bibinfo
  {author} {\bibfnamefont {F.}~\bibnamefont {Wiesner}}, \bibinfo {author}
  {\bibfnamefont {N.}~\bibnamefont {Walk}}, \bibinfo {author} {\bibfnamefont
  {J.}~\bibnamefont {Eisert}},\ and\ \bibinfo {author} {\bibfnamefont
  {J.}~\bibnamefont {Wolters}},\ }\href
  {https://doi.org/\url{10.1038/s42005-022-00945-9}} {\bibfield  {journal}
  {\bibinfo  {journal} {{Communications Physics}}\ }\textbf {\bibinfo {volume}
  {5}},\ \bibinfo {pages} {169} (\bibinfo {year} {2022})}\BibitemShut {NoStop}%
\bibitem [{\citenamefont {Liorni}\ \emph {et~al.}(2021)\citenamefont {Liorni},
  \citenamefont {Kampermann},\ and\ \citenamefont {Bru{\ss}}}]{Liorni.2021}%
  \BibitemOpen
  \bibfield  {author} {\bibinfo {author} {\bibfnamefont {C.}~\bibnamefont
  {Liorni}}, \bibinfo {author} {\bibfnamefont {H.}~\bibnamefont {Kampermann}},\
  and\ \bibinfo {author} {\bibfnamefont {D.}~\bibnamefont {Bru{\ss}}},\ }\href
  {https://doi.org/\url{10.1088/1367-2630/abfa63}} {\bibfield  {journal}
  {\bibinfo  {journal} {New J. Phys.}\ }\textbf {\bibinfo {volume} {23}},\
  \bibinfo {pages} {053021} (\bibinfo {year} {2021})}\BibitemShut {NoStop}%
\bibitem [{\citenamefont {Lvovsky}\ \emph {et~al.}(2009)\citenamefont
  {Lvovsky}, \citenamefont {Sanders},\ and\ \citenamefont
  {Tittel}}]{Lvovsky.2009}%
  \BibitemOpen
  \bibfield  {author} {\bibinfo {author} {\bibfnamefont {A.~I.}\ \bibnamefont
  {Lvovsky}}, \bibinfo {author} {\bibfnamefont {B.~C.}\ \bibnamefont
  {Sanders}},\ and\ \bibinfo {author} {\bibfnamefont {W.}~\bibnamefont
  {Tittel}},\ }\href {https://doi.org/\url{10.1038/nphoton.2009.231}}
  {\bibfield  {journal} {\bibinfo  {journal} {Nat. Photonics}\ }\textbf
  {\bibinfo {volume} {3}},\ \bibinfo {pages} {706} (\bibinfo {year}
  {2009})}\BibitemShut {NoStop}%
\bibitem [{\citenamefont {Heshami}\ \emph {et~al.}(2016)\citenamefont
  {Heshami}, \citenamefont {England}, \citenamefont {Humphreys}, \citenamefont
  {Bustard}, \citenamefont {Acosta}, \citenamefont {Nunn},\ and\ \citenamefont
  {Sussman}}]{Heshami.2016}%
  \BibitemOpen
  \bibfield  {author} {\bibinfo {author} {\bibfnamefont {K.}~\bibnamefont
  {Heshami}}, \bibinfo {author} {\bibfnamefont {D.~G.}\ \bibnamefont
  {England}}, \bibinfo {author} {\bibfnamefont {P.~C.}\ \bibnamefont
  {Humphreys}}, \bibinfo {author} {\bibfnamefont {P.~J.}\ \bibnamefont
  {Bustard}}, \bibinfo {author} {\bibfnamefont {V.~M.}\ \bibnamefont {Acosta}},
  \bibinfo {author} {\bibfnamefont {J.}~\bibnamefont {Nunn}},\ and\ \bibinfo
  {author} {\bibfnamefont {B.~J.}\ \bibnamefont {Sussman}},\ }\href
  {https://doi.org/\url{10.1080/09500340.2016.1148212}} {\bibfield  {journal}
  {\bibinfo  {journal} {J. Mod. Opt.}\ }\textbf {\bibinfo {volume} {63}},\
  \bibinfo {pages} {2005} (\bibinfo {year} {2016})}\BibitemShut {NoStop}%
\bibitem [{\citenamefont {Bao}\ \emph {et~al.}(2012)\citenamefont {Bao},
  \citenamefont {Reingruber}, \citenamefont {Dietrich}, \citenamefont {Rui},
  \citenamefont {D{\"u}ck}, \citenamefont {Strassel}, \citenamefont {Li},
  \citenamefont {Liu}, \citenamefont {Zhao},\ and\ \citenamefont
  {Pan}}]{Bao.2012}%
  \BibitemOpen
  \bibfield  {author} {\bibinfo {author} {\bibfnamefont {X.-H.}\ \bibnamefont
  {Bao}}, \bibinfo {author} {\bibfnamefont {A.}~\bibnamefont {Reingruber}},
  \bibinfo {author} {\bibfnamefont {P.}~\bibnamefont {Dietrich}}, \bibinfo
  {author} {\bibfnamefont {J.}~\bibnamefont {Rui}}, \bibinfo {author}
  {\bibfnamefont {A.}~\bibnamefont {D{\"u}ck}}, \bibinfo {author}
  {\bibfnamefont {T.}~\bibnamefont {Strassel}}, \bibinfo {author}
  {\bibfnamefont {L.}~\bibnamefont {Li}}, \bibinfo {author} {\bibfnamefont
  {N.-L.}\ \bibnamefont {Liu}}, \bibinfo {author} {\bibfnamefont
  {B.}~\bibnamefont {Zhao}},\ and\ \bibinfo {author} {\bibfnamefont {J.-W.}\
  \bibnamefont {Pan}},\ }\href {https://doi.org/\url{10.1038/nphys2324}}
  {\bibfield  {journal} {\bibinfo  {journal} {Nat. Phys.}\ }\textbf {\bibinfo
  {volume} {8}},\ \bibinfo {pages} {517} (\bibinfo {year} {2012})}\BibitemShut
  {NoStop}%
\bibitem [{\citenamefont {Yang}\ \emph {et~al.}(2016)\citenamefont {Yang},
  \citenamefont {Wang}, \citenamefont {Bao},\ and\ \citenamefont
  {Pan}}]{YangSJ.2016}%
  \BibitemOpen
  \bibfield  {author} {\bibinfo {author} {\bibfnamefont {S.-J.}\ \bibnamefont
  {Yang}}, \bibinfo {author} {\bibfnamefont {X.-J.}\ \bibnamefont {Wang}},
  \bibinfo {author} {\bibfnamefont {X.-H.}\ \bibnamefont {Bao}},\ and\ \bibinfo
  {author} {\bibfnamefont {J.-W.}\ \bibnamefont {Pan}},\ }\href
  {https://doi.org/\url{10.1038/nphoton.2016.51}} {\bibfield  {journal}
  {\bibinfo  {journal} {Nat. Photonics}\ }\textbf {\bibinfo {volume} {10}},\
  \bibinfo {pages} {381} (\bibinfo {year} {2016})}\BibitemShut {NoStop}%
\bibitem [{\citenamefont {Pu}\ \emph {et~al.}(2017)\citenamefont {Pu},
  \citenamefont {Jiang}, \citenamefont {Chang}, \citenamefont {Yang},
  \citenamefont {Li},\ and\ \citenamefont {Duan}}]{Pu.2017}%
  \BibitemOpen
  \bibfield  {author} {\bibinfo {author} {\bibfnamefont {Y.-F.}\ \bibnamefont
  {Pu}}, \bibinfo {author} {\bibfnamefont {N.}~\bibnamefont {Jiang}}, \bibinfo
  {author} {\bibfnamefont {W.}~\bibnamefont {Chang}}, \bibinfo {author}
  {\bibfnamefont {H.-X.}\ \bibnamefont {Yang}}, \bibinfo {author}
  {\bibfnamefont {C.}~\bibnamefont {Li}},\ and\ \bibinfo {author}
  {\bibfnamefont {L.-M.}\ \bibnamefont {Duan}},\ }\href
  {https://doi.org/\url{10.1038/ncomms15359}} {\bibfield  {journal} {\bibinfo
  {journal} {Nat. Commun.}\ }\textbf {\bibinfo {volume} {8}},\ \bibinfo {pages}
  {15359} (\bibinfo {year} {2017})}\BibitemShut {NoStop}%
\bibitem [{\citenamefont {Afzelius}\ \emph {et~al.}(2009)\citenamefont
  {Afzelius}, \citenamefont {Simon}, \citenamefont {Riedmatten},\ and\
  \citenamefont {Gisin}}]{Afzelius.2009}%
  \BibitemOpen
  \bibfield  {author} {\bibinfo {author} {\bibfnamefont {M.}~\bibnamefont
  {Afzelius}}, \bibinfo {author} {\bibfnamefont {C.}~\bibnamefont {Simon}},
  \bibinfo {author} {\bibfnamefont {H.~d.}\ \bibnamefont {Riedmatten}},\ and\
  \bibinfo {author} {\bibfnamefont {N.}~\bibnamefont {Gisin}},\ }\href
  {\url{10.1103/PhysRevA.79.052329}} {\bibfield  {journal} {\bibinfo  {journal}
  {Phys. Rev. A}\ }\textbf {\bibinfo {volume} {79}},\ \bibinfo {pages} {052329}
  (\bibinfo {year} {2009})}\BibitemShut {NoStop}%
\bibitem [{\citenamefont {G{\"u}ndogan}\ \emph {et~al.}(2015)\citenamefont
  {G{\"u}ndogan}, \citenamefont {Ledingham}, \citenamefont {Kutluer},
  \citenamefont {Mazzera},\ and\ \citenamefont
  {de~Riedmatten}}]{Gundogan.2015}%
  \BibitemOpen
  \bibfield  {author} {\bibinfo {author} {\bibfnamefont {M.}~\bibnamefont
  {G{\"u}ndogan}}, \bibinfo {author} {\bibfnamefont {P.~M.}\ \bibnamefont
  {Ledingham}}, \bibinfo {author} {\bibfnamefont {K.}~\bibnamefont {Kutluer}},
  \bibinfo {author} {\bibfnamefont {M.}~\bibnamefont {Mazzera}},\ and\ \bibinfo
  {author} {\bibfnamefont {H.}~\bibnamefont {de~Riedmatten}},\ }\href
  {https://doi.org/\url{10.1103/PhysRevLett.114.230501}} {\bibfield  {journal}
  {\bibinfo  {journal} {Phys. Rev. Lett.}\ }\textbf {\bibinfo {volume} {114}},\
  \bibinfo {pages} {230501} (\bibinfo {year} {2015})}\BibitemShut {NoStop}%
\bibitem [{\citenamefont {Yang}\ \emph {et~al.}(2018)\citenamefont {Yang},
  \citenamefont {Zhou}, \citenamefont {Hua}, \citenamefont {Liu}, \citenamefont
  {Li}, \citenamefont {Li}, \citenamefont {Ma}, \citenamefont {Liu},
  \citenamefont {Liang}, \citenamefont {Li}, \citenamefont {Xiao},
  \citenamefont {Hu}, \citenamefont {Li},\ and\ \citenamefont
  {Guo}}]{YangTS.2018}%
  \BibitemOpen
  \bibfield  {author} {\bibinfo {author} {\bibfnamefont {T.-S.}\ \bibnamefont
  {Yang}}, \bibinfo {author} {\bibfnamefont {Z.-Q.}\ \bibnamefont {Zhou}},
  \bibinfo {author} {\bibfnamefont {Y.-L.}\ \bibnamefont {Hua}}, \bibinfo
  {author} {\bibfnamefont {X.}~\bibnamefont {Liu}}, \bibinfo {author}
  {\bibfnamefont {Z.-F.}\ \bibnamefont {Li}}, \bibinfo {author} {\bibfnamefont
  {P.-Y.}\ \bibnamefont {Li}}, \bibinfo {author} {\bibfnamefont
  {Y.}~\bibnamefont {Ma}}, \bibinfo {author} {\bibfnamefont {C.}~\bibnamefont
  {Liu}}, \bibinfo {author} {\bibfnamefont {P.-J.}\ \bibnamefont {Liang}},
  \bibinfo {author} {\bibfnamefont {X.}~\bibnamefont {Li}}, \bibinfo {author}
  {\bibfnamefont {Y.-X.}\ \bibnamefont {Xiao}}, \bibinfo {author}
  {\bibfnamefont {J.}~\bibnamefont {Hu}}, \bibinfo {author} {\bibfnamefont
  {C.-F.}\ \bibnamefont {Li}},\ and\ \bibinfo {author} {\bibfnamefont {G.-C.}\
  \bibnamefont {Guo}},\ }\href
  {https://doi.org/\url{10.1038/s41467-018-05669-5}} {\bibfield  {journal}
  {\bibinfo  {journal} {Nat. Commun.}\ }\textbf {\bibinfo {volume} {9}},\
  \bibinfo {pages} {3407} (\bibinfo {year} {2018})}\BibitemShut {NoStop}%
\bibitem [{\citenamefont {Ma}\ \emph {et~al.}(2021)\citenamefont {Ma},
  \citenamefont {Ma}, \citenamefont {Zhou}, \citenamefont {Li},\ and\
  \citenamefont {Guo}}]{Ma.2021}%
  \BibitemOpen
  \bibfield  {author} {\bibinfo {author} {\bibfnamefont {Y.}~\bibnamefont
  {Ma}}, \bibinfo {author} {\bibfnamefont {Y.-Z.}\ \bibnamefont {Ma}}, \bibinfo
  {author} {\bibfnamefont {Z.-Q.}\ \bibnamefont {Zhou}}, \bibinfo {author}
  {\bibfnamefont {C.-F.}\ \bibnamefont {Li}},\ and\ \bibinfo {author}
  {\bibfnamefont {G.-C.}\ \bibnamefont {Guo}},\ }\href
  {https://doi.org/\url{10.1038/s41467-021-22706-y}} {\bibfield  {journal}
  {\bibinfo  {journal} {Nat. Commun.}\ }\textbf {\bibinfo {volume} {12}},\
  \bibinfo {pages} {2381} (\bibinfo {year} {2021})}\BibitemShut {NoStop}%
\bibitem [{\citenamefont {Ortu}\ \emph {et~al.}(2022)\citenamefont {Ortu},
  \citenamefont {Holz{\"a}pfel}, \citenamefont {Etesse},\ and\ \citenamefont
  {Afzelius}}]{Ortu.2022}%
  \BibitemOpen
  \bibfield  {author} {\bibinfo {author} {\bibfnamefont {A.}~\bibnamefont
  {Ortu}}, \bibinfo {author} {\bibfnamefont {A.}~\bibnamefont {Holz{\"a}pfel}},
  \bibinfo {author} {\bibfnamefont {J.}~\bibnamefont {Etesse}},\ and\ \bibinfo
  {author} {\bibfnamefont {M.}~\bibnamefont {Afzelius}},\ }\href
  {https://doi.org/\url{10.1038/s41534-022-00541-3}} {\bibfield  {journal}
  {\bibinfo  {journal} {{npj Quantum Information}}\ }\textbf {\bibinfo {volume}
  {8}},\ \bibinfo {pages} {29} (\bibinfo {year} {2022})}\BibitemShut {NoStop}%
\bibitem [{\citenamefont {Saunders}\ \emph {et~al.}(2016)\citenamefont
  {Saunders}, \citenamefont {Munns}, \citenamefont {Champion}, \citenamefont
  {Qiu}, \citenamefont {Kaczmarek}, \citenamefont {Poem}, \citenamefont
  {Ledingham}, \citenamefont {Walmsley},\ and\ \citenamefont
  {Nunn}}]{Saunders.2016}%
  \BibitemOpen
  \bibfield  {author} {\bibinfo {author} {\bibfnamefont {D.~J.}\ \bibnamefont
  {Saunders}}, \bibinfo {author} {\bibfnamefont {J.~H.~D.}\ \bibnamefont
  {Munns}}, \bibinfo {author} {\bibfnamefont {T.~F.~M.}\ \bibnamefont
  {Champion}}, \bibinfo {author} {\bibfnamefont {C.}~\bibnamefont {Qiu}},
  \bibinfo {author} {\bibfnamefont {K.~T.}\ \bibnamefont {Kaczmarek}}, \bibinfo
  {author} {\bibfnamefont {E.}~\bibnamefont {Poem}}, \bibinfo {author}
  {\bibfnamefont {P.~M.}\ \bibnamefont {Ledingham}}, \bibinfo {author}
  {\bibfnamefont {I.~A.}\ \bibnamefont {Walmsley}},\ and\ \bibinfo {author}
  {\bibfnamefont {J.}~\bibnamefont {Nunn}},\ }\href
  {https://doi.org/\url{10.1103/PhysRevLett.116.090501}} {\bibfield  {journal}
  {\bibinfo  {journal} {Phys. Rev. Lett.}\ }\textbf {\bibinfo {volume} {116}},\
  \bibinfo {pages} {090501} (\bibinfo {year} {2016})}\BibitemShut {NoStop}%
\bibitem [{\citenamefont {Wolters}\ \emph {et~al.}(2017)\citenamefont
  {Wolters}, \citenamefont {Buser}, \citenamefont {Horsley}, \citenamefont
  {B{\'e}guin}, \citenamefont {J{\"o}ckel}, \citenamefont {Jahn}, \citenamefont
  {Warburton},\ and\ \citenamefont {Treutlein}}]{Wolters.2017}%
  \BibitemOpen
  \bibfield  {author} {\bibinfo {author} {\bibfnamefont {J.}~\bibnamefont
  {Wolters}}, \bibinfo {author} {\bibfnamefont {G.}~\bibnamefont {Buser}},
  \bibinfo {author} {\bibfnamefont {A.}~\bibnamefont {Horsley}}, \bibinfo
  {author} {\bibfnamefont {L.}~\bibnamefont {B{\'e}guin}}, \bibinfo {author}
  {\bibfnamefont {A.}~\bibnamefont {J{\"o}ckel}}, \bibinfo {author}
  {\bibfnamefont {J.~P.}\ \bibnamefont {Jahn}}, \bibinfo {author}
  {\bibfnamefont {R.~J.}\ \bibnamefont {Warburton}},\ and\ \bibinfo {author}
  {\bibfnamefont {P.}~\bibnamefont {Treutlein}},\ }\href
  {https://doi.org/\url{10.1103/PhysRevLett.119.060502}} {\bibfield  {journal}
  {\bibinfo  {journal} {Phys. Rev. Lett.}\ }\textbf {\bibinfo {volume} {119}},\
  \bibinfo {pages} {1} (\bibinfo {year} {2017})}\BibitemShut {NoStop}%
\bibitem [{\citenamefont {Katz}\ and\ \citenamefont
  {Firstenberg}(2018)}]{Katz.2018}%
  \BibitemOpen
  \bibfield  {author} {\bibinfo {author} {\bibfnamefont {O.}~\bibnamefont
  {Katz}}\ and\ \bibinfo {author} {\bibfnamefont {O.}~\bibnamefont
  {Firstenberg}},\ }\href {https://doi.org/\url{10.1038/s41467-018-04458-4}}
  {\bibfield  {journal} {\bibinfo  {journal} {Nat. Commun.}\ }\textbf {\bibinfo
  {volume} {9}},\ \bibinfo {pages} {2074} (\bibinfo {year} {2018})}\BibitemShut
  {NoStop}%
\bibitem [{\citenamefont {Guo}\ \emph {et~al.}(2019)\citenamefont {Guo},
  \citenamefont {Feng}, \citenamefont {Yang}, \citenamefont {Yu}, \citenamefont
  {Chen}, \citenamefont {Yuan},\ and\ \citenamefont {Zhang}}]{Guo.2019}%
  \BibitemOpen
  \bibfield  {author} {\bibinfo {author} {\bibfnamefont {J.}~\bibnamefont
  {Guo}}, \bibinfo {author} {\bibfnamefont {X.}~\bibnamefont {Feng}}, \bibinfo
  {author} {\bibfnamefont {P.}~\bibnamefont {Yang}}, \bibinfo {author}
  {\bibfnamefont {Z.}~\bibnamefont {Yu}}, \bibinfo {author} {\bibfnamefont
  {L.~Q.}\ \bibnamefont {Chen}}, \bibinfo {author} {\bibfnamefont {C.-h.}\
  \bibnamefont {Yuan}},\ and\ \bibinfo {author} {\bibfnamefont
  {W.}~\bibnamefont {Zhang}},\ }\href {\url{10.1038/s41467-018-08118-5}}
  {\bibfield  {journal} {\bibinfo  {journal} {Nat. Commun.}\ }\textbf {\bibinfo
  {volume} {10}},\ \bibinfo {pages} {148} (\bibinfo {year} {2019})}\BibitemShut
  {NoStop}%
\bibitem [{\citenamefont {Thomas}\ \emph {et~al.}(2019)\citenamefont {Thomas},
  \citenamefont {Hird}, \citenamefont {Munns}, \citenamefont {Brecht},
  \citenamefont {Saunders}, \citenamefont {Nunn}, \citenamefont {Walmsley},\
  and\ \citenamefont {Ledingham}}]{Thomas.2019}%
  \BibitemOpen
  \bibfield  {author} {\bibinfo {author} {\bibfnamefont {S.~E.}\ \bibnamefont
  {Thomas}}, \bibinfo {author} {\bibfnamefont {T.~M.}\ \bibnamefont {Hird}},
  \bibinfo {author} {\bibfnamefont {J.~H.~D.}\ \bibnamefont {Munns}}, \bibinfo
  {author} {\bibfnamefont {B.}~\bibnamefont {Brecht}}, \bibinfo {author}
  {\bibfnamefont {D.~J.}\ \bibnamefont {Saunders}}, \bibinfo {author}
  {\bibfnamefont {J.}~\bibnamefont {Nunn}}, \bibinfo {author} {\bibfnamefont
  {I.~A.}\ \bibnamefont {Walmsley}},\ and\ \bibinfo {author} {\bibfnamefont
  {P.~M.}\ \bibnamefont {Ledingham}},\ }\href
  {https://doi.org/\url{10.1103/PhysRevA.100.033801}} {\bibfield  {journal}
  {\bibinfo  {journal} {Phys. Rev. A}\ }\textbf {\bibinfo {volume} {100}},\
  \bibinfo {pages} {033801} (\bibinfo {year} {2019})}\BibitemShut {NoStop}%
\bibitem [{\citenamefont {Michelberger}\ \emph {et~al.}(2015)\citenamefont
  {Michelberger}, \citenamefont {Champion}, \citenamefont {Sprague},
  \citenamefont {Kaczmarek}, \citenamefont {Barbieri}, \citenamefont {Jin},
  \citenamefont {England}, \citenamefont {Kolthammer}, \citenamefont
  {Saunders}, \citenamefont {Nunn},\ and\ \citenamefont
  {Walmsley}}]{Michelberger.2015}%
  \BibitemOpen
  \bibfield  {author} {\bibinfo {author} {\bibfnamefont {P.~S.}\ \bibnamefont
  {Michelberger}}, \bibinfo {author} {\bibfnamefont {T.~F.~M.}\ \bibnamefont
  {Champion}}, \bibinfo {author} {\bibfnamefont {M.~R.}\ \bibnamefont
  {Sprague}}, \bibinfo {author} {\bibfnamefont {K.~T.}\ \bibnamefont
  {Kaczmarek}}, \bibinfo {author} {\bibfnamefont {M.}~\bibnamefont {Barbieri}},
  \bibinfo {author} {\bibfnamefont {X.~M.}\ \bibnamefont {Jin}}, \bibinfo
  {author} {\bibfnamefont {D.~G.}\ \bibnamefont {England}}, \bibinfo {author}
  {\bibfnamefont {W.~S.}\ \bibnamefont {Kolthammer}}, \bibinfo {author}
  {\bibfnamefont {D.~J.}\ \bibnamefont {Saunders}}, \bibinfo {author}
  {\bibfnamefont {J.}~\bibnamefont {Nunn}},\ and\ \bibinfo {author}
  {\bibfnamefont {I.~A.}\ \bibnamefont {Walmsley}},\ }\href
  {https://doi.org/\url{10.1088/1367-2630/17/4/043006}} {\bibfield  {journal}
  {\bibinfo  {journal} {New J. Phys.}\ }\textbf {\bibinfo {volume} {17}},\
  \bibinfo {pages} {043006} (\bibinfo {year} {2015})}\BibitemShut {NoStop}%
\bibitem [{\citenamefont {Kimble}(2008)}]{Kimble.2008}%
  \BibitemOpen
  \bibfield  {author} {\bibinfo {author} {\bibfnamefont {H.~J.}\ \bibnamefont
  {Kimble}},\ }\href {https://doi.org/\url{10.1038/nature07127}} {\bibfield
  {journal} {\bibinfo  {journal} {Nature}\ }\textbf {\bibinfo {volume} {453}},\
  \bibinfo {pages} {1023} (\bibinfo {year} {2008})}\BibitemShut {NoStop}%
\bibitem [{\citenamefont {Tang}\ \emph {et~al.}(2015)\citenamefont {Tang},
  \citenamefont {Zhou}, \citenamefont {Wang}, \citenamefont {Li}, \citenamefont
  {Liu}, \citenamefont {Hua}, \citenamefont {Zou}, \citenamefont {Wang},
  \citenamefont {He}, \citenamefont {Chen}, \citenamefont {Sun}, \citenamefont
  {Yu}, \citenamefont {Li}, \citenamefont {Zha}, \citenamefont {Ni},
  \citenamefont {Niu}, \citenamefont {Li},\ and\ \citenamefont
  {Guo}}]{Tang.2015}%
  \BibitemOpen
  \bibfield  {author} {\bibinfo {author} {\bibfnamefont {J.-S.}\ \bibnamefont
  {Tang}}, \bibinfo {author} {\bibfnamefont {Z.-Q.}\ \bibnamefont {Zhou}},
  \bibinfo {author} {\bibfnamefont {Y.-T.}\ \bibnamefont {Wang}}, \bibinfo
  {author} {\bibfnamefont {Y.-L.}\ \bibnamefont {Li}}, \bibinfo {author}
  {\bibfnamefont {X.}~\bibnamefont {Liu}}, \bibinfo {author} {\bibfnamefont
  {Y.-L.}\ \bibnamefont {Hua}}, \bibinfo {author} {\bibfnamefont
  {Y.}~\bibnamefont {Zou}}, \bibinfo {author} {\bibfnamefont {S.}~\bibnamefont
  {Wang}}, \bibinfo {author} {\bibfnamefont {D.-Y.}\ \bibnamefont {He}},
  \bibinfo {author} {\bibfnamefont {G.}~\bibnamefont {Chen}}, \bibinfo {author}
  {\bibfnamefont {Y.-N.}\ \bibnamefont {Sun}}, \bibinfo {author} {\bibfnamefont
  {Y.}~\bibnamefont {Yu}}, \bibinfo {author} {\bibfnamefont {M.-F.}\
  \bibnamefont {Li}}, \bibinfo {author} {\bibfnamefont {G.-W.}\ \bibnamefont
  {Zha}}, \bibinfo {author} {\bibfnamefont {H.-Q.}\ \bibnamefont {Ni}},
  \bibinfo {author} {\bibfnamefont {Z.-C.}\ \bibnamefont {Niu}}, \bibinfo
  {author} {\bibfnamefont {C.-F.}\ \bibnamefont {Li}},\ and\ \bibinfo {author}
  {\bibfnamefont {G.-C.}\ \bibnamefont {Guo}},\ }\href
  {https://doi.org/\url{10.1038/ncomms9652}} {\bibfield  {journal} {\bibinfo
  {journal} {Nat. Commun.}\ }\textbf {\bibinfo {volume} {6}},\ \bibinfo {pages}
  {8652} (\bibinfo {year} {2015})}\BibitemShut {NoStop}%
\bibitem [{\citenamefont {Somaschi}\ \emph {et~al.}(2016)\citenamefont
  {Somaschi}, \citenamefont {Giesz}, \citenamefont {de~Santis}, \citenamefont
  {Loredo}, \citenamefont {Almeida}, \citenamefont {Hornecker}, \citenamefont
  {Portalupi}, \citenamefont {Grange}, \citenamefont {Ant{\'o}n}, \citenamefont
  {Demory}, \citenamefont {G{\'o}mez}, \citenamefont {Sagnes}, \citenamefont
  {Lanzillotti-Kimura}, \citenamefont {Lema{\'i}tre}, \citenamefont {Auffeves},
  \citenamefont {White}, \citenamefont {Lanco},\ and\ \citenamefont
  {Senellart}}]{Somaschi.2016}%
  \BibitemOpen
  \bibfield  {author} {\bibinfo {author} {\bibfnamefont {N.}~\bibnamefont
  {Somaschi}}, \bibinfo {author} {\bibfnamefont {V.}~\bibnamefont {Giesz}},
  \bibinfo {author} {\bibfnamefont {L.}~\bibnamefont {de~Santis}}, \bibinfo
  {author} {\bibfnamefont {J.~C.}\ \bibnamefont {Loredo}}, \bibinfo {author}
  {\bibfnamefont {M.~P.}\ \bibnamefont {Almeida}}, \bibinfo {author}
  {\bibfnamefont {G.}~\bibnamefont {Hornecker}}, \bibinfo {author}
  {\bibfnamefont {S.~L.}\ \bibnamefont {Portalupi}}, \bibinfo {author}
  {\bibfnamefont {T.}~\bibnamefont {Grange}}, \bibinfo {author} {\bibfnamefont
  {C.}~\bibnamefont {Ant{\'o}n}}, \bibinfo {author} {\bibfnamefont
  {J.}~\bibnamefont {Demory}}, \bibinfo {author} {\bibfnamefont
  {C.}~\bibnamefont {G{\'o}mez}}, \bibinfo {author} {\bibfnamefont
  {I.}~\bibnamefont {Sagnes}}, \bibinfo {author} {\bibfnamefont {N.~D.}\
  \bibnamefont {Lanzillotti-Kimura}}, \bibinfo {author} {\bibfnamefont
  {A.}~\bibnamefont {Lema{\'i}tre}}, \bibinfo {author} {\bibfnamefont
  {A.}~\bibnamefont {Auffeves}}, \bibinfo {author} {\bibfnamefont {A.~G.}\
  \bibnamefont {White}}, \bibinfo {author} {\bibfnamefont {L.}~\bibnamefont
  {Lanco}},\ and\ \bibinfo {author} {\bibfnamefont {P.}~\bibnamefont
  {Senellart}},\ }\href {https://doi.org/\url{10.1038/nphoton.2016.23}}
  {\bibfield  {journal} {\bibinfo  {journal} {Nat. Photonics}\ }\textbf
  {\bibinfo {volume} {10}},\ \bibinfo {pages} {340} (\bibinfo {year}
  {2016})}\BibitemShut {NoStop}%
\bibitem [{\citenamefont {Ding}\ \emph {et~al.}(2016)\citenamefont {Ding},
  \citenamefont {He}, \citenamefont {Duan}, \citenamefont {Gregersen},
  \citenamefont {Chen}, \citenamefont {Unsleber}, \citenamefont {Maier},
  \citenamefont {Schneider}, \citenamefont {Kamp}, \citenamefont {H{\"o}fling},
  \citenamefont {Lu},\ and\ \citenamefont {Pan}}]{Ding.2016}%
  \BibitemOpen
  \bibfield  {author} {\bibinfo {author} {\bibfnamefont {X.}~\bibnamefont
  {Ding}}, \bibinfo {author} {\bibfnamefont {Y.}~\bibnamefont {He}}, \bibinfo
  {author} {\bibfnamefont {Z.~C.}\ \bibnamefont {Duan}}, \bibinfo {author}
  {\bibfnamefont {N.}~\bibnamefont {Gregersen}}, \bibinfo {author}
  {\bibfnamefont {M.~C.}\ \bibnamefont {Chen}}, \bibinfo {author}
  {\bibfnamefont {S.}~\bibnamefont {Unsleber}}, \bibinfo {author}
  {\bibfnamefont {S.}~\bibnamefont {Maier}}, \bibinfo {author} {\bibfnamefont
  {C.}~\bibnamefont {Schneider}}, \bibinfo {author} {\bibfnamefont
  {M.}~\bibnamefont {Kamp}}, \bibinfo {author} {\bibfnamefont {S.}~\bibnamefont
  {H{\"o}fling}}, \bibinfo {author} {\bibfnamefont {C.~Y.}\ \bibnamefont
  {Lu}},\ and\ \bibinfo {author} {\bibfnamefont {J.~W.}\ \bibnamefont {Pan}},\
  }\href {https://doi.org/\url{10.1103/PhysRevLett.116.020401}} {\bibfield
  {journal} {\bibinfo  {journal} {Phys. Rev. Lett.}\ }\textbf {\bibinfo
  {volume} {116}},\ \bibinfo {pages} {020401} (\bibinfo {year}
  {2016})}\BibitemShut {NoStop}%
\bibitem [{\citenamefont {B{\'e}guin}\ \emph {et~al.}(2018)\citenamefont
  {B{\'e}guin}, \citenamefont {Jahn}, \citenamefont {Wolters}, \citenamefont
  {Reindl}, \citenamefont {Huo}, \citenamefont {Trotta}, \citenamefont
  {Rastelli}, \citenamefont {Ding}, \citenamefont {Schmidt}, \citenamefont
  {Treutlein},\ and\ \citenamefont {Warburton}}]{Beguin.2018}%
  \BibitemOpen
  \bibfield  {author} {\bibinfo {author} {\bibfnamefont {L.}~\bibnamefont
  {B{\'e}guin}}, \bibinfo {author} {\bibfnamefont {J.-P.}\ \bibnamefont
  {Jahn}}, \bibinfo {author} {\bibfnamefont {J.}~\bibnamefont {Wolters}},
  \bibinfo {author} {\bibfnamefont {M.}~\bibnamefont {Reindl}}, \bibinfo
  {author} {\bibfnamefont {Y.}~\bibnamefont {Huo}}, \bibinfo {author}
  {\bibfnamefont {R.}~\bibnamefont {Trotta}}, \bibinfo {author} {\bibfnamefont
  {A.}~\bibnamefont {Rastelli}}, \bibinfo {author} {\bibfnamefont
  {F.}~\bibnamefont {Ding}}, \bibinfo {author} {\bibfnamefont {O.~G.}\
  \bibnamefont {Schmidt}}, \bibinfo {author} {\bibfnamefont {P.}~\bibnamefont
  {Treutlein}},\ and\ \bibinfo {author} {\bibfnamefont {R.~J.}\ \bibnamefont
  {Warburton}},\ }\href {https://doi.org/\url{10.1103/PhysRevB.97.205304}}
  {\bibfield  {journal} {\bibinfo  {journal} {Phys. Rev. B}\ }\textbf {\bibinfo
  {volume} {97}},\ \bibinfo {pages} {205304} (\bibinfo {year}
  {2018})}\BibitemShut {NoStop}%
\bibitem [{\citenamefont {Kroh}\ \emph {et~al.}(2019)\citenamefont {Kroh},
  \citenamefont {Wolters}, \citenamefont {Ahlrichs}, \citenamefont {Schell},
  \citenamefont {Thoma}, \citenamefont {Reitzenstein}, \citenamefont
  {Wildmann}, \citenamefont {Zallo}, \citenamefont {Trotta}, \citenamefont
  {Rastelli}, \citenamefont {Schmidt},\ and\ \citenamefont
  {Benson}}]{Kroh.2019}%
  \BibitemOpen
  \bibfield  {author} {\bibinfo {author} {\bibfnamefont {T.}~\bibnamefont
  {Kroh}}, \bibinfo {author} {\bibfnamefont {J.}~\bibnamefont {Wolters}},
  \bibinfo {author} {\bibfnamefont {A.}~\bibnamefont {Ahlrichs}}, \bibinfo
  {author} {\bibfnamefont {A.~W.}\ \bibnamefont {Schell}}, \bibinfo {author}
  {\bibfnamefont {A.}~\bibnamefont {Thoma}}, \bibinfo {author} {\bibfnamefont
  {S.}~\bibnamefont {Reitzenstein}}, \bibinfo {author} {\bibfnamefont {J.~S.}\
  \bibnamefont {Wildmann}}, \bibinfo {author} {\bibfnamefont {E.}~\bibnamefont
  {Zallo}}, \bibinfo {author} {\bibfnamefont {R.}~\bibnamefont {Trotta}},
  \bibinfo {author} {\bibfnamefont {A.}~\bibnamefont {Rastelli}}, \bibinfo
  {author} {\bibfnamefont {O.~G.}\ \bibnamefont {Schmidt}},\ and\ \bibinfo
  {author} {\bibfnamefont {O.}~\bibnamefont {Benson}},\ }\href
  {https://doi.org/\url{10.1038/s41598-019-50062-x}} {\bibfield  {journal}
  {\bibinfo  {journal} {{Scientific Reports}}\ }\textbf {\bibinfo {volume}
  {9}},\ \bibinfo {pages} {13728} (\bibinfo {year} {2019})}\BibitemShut
  {NoStop}%
\bibitem [{\citenamefont {Schunk}\ \emph {et~al.}(2015)\citenamefont {Schunk},
  \citenamefont {Vogl}, \citenamefont {Strekalov}, \citenamefont {F{\"o}rtsch},
  \citenamefont {Sedlmeir}, \citenamefont {Schwefel}, \citenamefont
  {G{\"o}belt}, \citenamefont {Christiansen}, \citenamefont {Leuchs},\ and\
  \citenamefont {Marquardt}}]{Schunk.2015}%
  \BibitemOpen
  \bibfield  {author} {\bibinfo {author} {\bibfnamefont {G.}~\bibnamefont
  {Schunk}}, \bibinfo {author} {\bibfnamefont {U.}~\bibnamefont {Vogl}},
  \bibinfo {author} {\bibfnamefont {D.~V.}\ \bibnamefont {Strekalov}}, \bibinfo
  {author} {\bibfnamefont {M.}~\bibnamefont {F{\"o}rtsch}}, \bibinfo {author}
  {\bibfnamefont {F.}~\bibnamefont {Sedlmeir}}, \bibinfo {author}
  {\bibfnamefont {H.~G.~L.}\ \bibnamefont {Schwefel}}, \bibinfo {author}
  {\bibfnamefont {M.}~\bibnamefont {G{\"o}belt}}, \bibinfo {author}
  {\bibfnamefont {S.}~\bibnamefont {Christiansen}}, \bibinfo {author}
  {\bibfnamefont {G.}~\bibnamefont {Leuchs}},\ and\ \bibinfo {author}
  {\bibfnamefont {C.}~\bibnamefont {Marquardt}},\ }\href
  {https://doi.org/\url{10.1364/OPTICA.2.000773}} {\bibfield  {journal}
  {\bibinfo  {journal} {Optica}\ }\textbf {\bibinfo {volume} {2}},\ \bibinfo
  {pages} {773} (\bibinfo {year} {2015})}\BibitemShut {NoStop}%
\bibitem [{\citenamefont {Seri}\ \emph {et~al.}(2017)\citenamefont {Seri},
  \citenamefont {Lenhard}, \citenamefont {Riel{\"a}nder}, \citenamefont
  {G{\"u}ndogan}, \citenamefont {Ledingham}, \citenamefont {Mazzera},\ and\
  \citenamefont {de~Riedmatten}}]{Seri.2017}%
  \BibitemOpen
  \bibfield  {author} {\bibinfo {author} {\bibfnamefont {A.}~\bibnamefont
  {Seri}}, \bibinfo {author} {\bibfnamefont {A.}~\bibnamefont {Lenhard}},
  \bibinfo {author} {\bibfnamefont {D.}~\bibnamefont {Riel{\"a}nder}}, \bibinfo
  {author} {\bibfnamefont {M.}~\bibnamefont {G{\"u}ndogan}}, \bibinfo {author}
  {\bibfnamefont {P.~M.}\ \bibnamefont {Ledingham}}, \bibinfo {author}
  {\bibfnamefont {M.}~\bibnamefont {Mazzera}},\ and\ \bibinfo {author}
  {\bibfnamefont {H.}~\bibnamefont {de~Riedmatten}},\ }\href
  {https://doi.org/\url{10.1103/PhysRevX.7.021028}} {\bibfield  {journal}
  {\bibinfo  {journal} {Phys. Rev. X}\ }\textbf {\bibinfo {volume} {7}},\
  \bibinfo {pages} {021028} (\bibinfo {year} {2017})}\BibitemShut {NoStop}%
\bibitem [{\citenamefont {Mottola}\ \emph {et~al.}(2020)\citenamefont
  {Mottola}, \citenamefont {Buser}, \citenamefont {M{\"u}ller}, \citenamefont
  {Kroh}, \citenamefont {Ahlrichs}, \citenamefont {Ramelow}, \citenamefont
  {Benson}, \citenamefont {Treutlein},\ and\ \citenamefont
  {Wolters}}]{Mottola.2020}%
  \BibitemOpen
  \bibfield  {author} {\bibinfo {author} {\bibfnamefont {R.}~\bibnamefont
  {Mottola}}, \bibinfo {author} {\bibfnamefont {G.}~\bibnamefont {Buser}},
  \bibinfo {author} {\bibfnamefont {C.}~\bibnamefont {M{\"u}ller}}, \bibinfo
  {author} {\bibfnamefont {T.}~\bibnamefont {Kroh}}, \bibinfo {author}
  {\bibfnamefont {A.}~\bibnamefont {Ahlrichs}}, \bibinfo {author}
  {\bibfnamefont {S.}~\bibnamefont {Ramelow}}, \bibinfo {author} {\bibfnamefont
  {O.}~\bibnamefont {Benson}}, \bibinfo {author} {\bibfnamefont
  {P.}~\bibnamefont {Treutlein}},\ and\ \bibinfo {author} {\bibfnamefont
  {J.}~\bibnamefont {Wolters}},\ }\href
  {https://doi.org/\url{10.1364/oe.384081}} {\bibfield  {journal} {\bibinfo
  {journal} {Opt. Express}\ }\textbf {\bibinfo {volume} {28}},\ \bibinfo
  {pages} {3159} (\bibinfo {year} {2020})}\BibitemShut {NoStop}%
\bibitem [{\citenamefont {Buser}\ \emph {et~al.}(2022)\citenamefont {Buser},
  \citenamefont {Mottola}, \citenamefont {Cotting}, \citenamefont {Wolters},\
  and\ \citenamefont {Treutlein}}]{Buser.2022}%
  \BibitemOpen
  \bibfield  {author} {\bibinfo {author} {\bibfnamefont {G.}~\bibnamefont
  {Buser}}, \bibinfo {author} {\bibfnamefont {R.}~\bibnamefont {Mottola}},
  \bibinfo {author} {\bibfnamefont {B.}~\bibnamefont {Cotting}}, \bibinfo
  {author} {\bibfnamefont {J.}~\bibnamefont {Wolters}},\ and\ \bibinfo {author}
  {\bibfnamefont {P.}~\bibnamefont {Treutlein}},\ }\href
  {https://doi.org/\url{10.1103/PRXQuantum.3.020349}} {\bibfield  {journal}
  {\bibinfo  {journal} {{PRX Quantum}}\ }\textbf {\bibinfo {volume} {3}},\
  \bibinfo {pages} {020349} (\bibinfo {year} {2022})}\BibitemShut {NoStop}%
\bibitem [{\citenamefont {Dideriksen}\ \emph {et~al.}(2021)\citenamefont
  {Dideriksen}, \citenamefont {Schmieg}, \citenamefont {Zugenmaier},\ and\
  \citenamefont {Polzik}}]{Dideriksen.2021}%
  \BibitemOpen
  \bibfield  {author} {\bibinfo {author} {\bibfnamefont {K.~B.}\ \bibnamefont
  {Dideriksen}}, \bibinfo {author} {\bibfnamefont {R.}~\bibnamefont {Schmieg}},
  \bibinfo {author} {\bibfnamefont {M.}~\bibnamefont {Zugenmaier}},\ and\
  \bibinfo {author} {\bibfnamefont {E.~S.}\ \bibnamefont {Polzik}},\ }\href
  {https://doi.org/\url{10.1038/s41467-021-24033-8}} {\bibfield  {journal}
  {\bibinfo  {journal} {Nat. Commun.}\ }\textbf {\bibinfo {volume} {12}},\
  \bibinfo {pages} {3699} (\bibinfo {year} {2021})}\BibitemShut {NoStop}%
\bibitem [{\citenamefont {Davidson}\ \emph {et~al.}(2021)\citenamefont
  {Davidson}, \citenamefont {Finkelstein}, \citenamefont {Poem},\ and\
  \citenamefont {Firstenberg}}]{Davidson.2021}%
  \BibitemOpen
  \bibfield  {author} {\bibinfo {author} {\bibfnamefont {O.}~\bibnamefont
  {Davidson}}, \bibinfo {author} {\bibfnamefont {R.}~\bibnamefont
  {Finkelstein}}, \bibinfo {author} {\bibfnamefont {E.}~\bibnamefont {Poem}},\
  and\ \bibinfo {author} {\bibfnamefont {O.}~\bibnamefont {Firstenberg}},\
  }\href {https://doi.org/\url{10.1088/1367-2630/ac14ab}} {\bibfield  {journal}
  {\bibinfo  {journal} {New J. Phys.}\ }\textbf {\bibinfo {volume} {23}},\
  \bibinfo {pages} {073050} (\bibinfo {year} {2021})}\BibitemShut {NoStop}%
\bibitem [{\citenamefont {Jobez}\ \emph {et~al.}(2015)\citenamefont {Jobez},
  \citenamefont {Laplane}, \citenamefont {Timoney}, \citenamefont {Gisin},
  \citenamefont {Ferrier}, \citenamefont {Goldner},\ and\ \citenamefont
  {Afzelius}}]{Jobez.2015}%
  \BibitemOpen
  \bibfield  {author} {\bibinfo {author} {\bibfnamefont {P.}~\bibnamefont
  {Jobez}}, \bibinfo {author} {\bibfnamefont {C.}~\bibnamefont {Laplane}},
  \bibinfo {author} {\bibfnamefont {N.}~\bibnamefont {Timoney}}, \bibinfo
  {author} {\bibfnamefont {N.}~\bibnamefont {Gisin}}, \bibinfo {author}
  {\bibfnamefont {A.}~\bibnamefont {Ferrier}}, \bibinfo {author} {\bibfnamefont
  {P.}~\bibnamefont {Goldner}},\ and\ \bibinfo {author} {\bibfnamefont
  {M.}~\bibnamefont {Afzelius}},\ }\href
  {https://doi.org/\url{10.1103/PhysRevLett.114.230502}} {\bibfield  {journal}
  {\bibinfo  {journal} {Phys. Rev. Lett.}\ }\textbf {\bibinfo {volume} {114}},\
  \bibinfo {pages} {230502} (\bibinfo {year} {2015})}\BibitemShut {NoStop}%
\bibitem [{\citenamefont {Fleischhauer}\ \emph {et~al.}(2005)\citenamefont
  {Fleischhauer}, \citenamefont {Imamoglu},\ and\ \citenamefont
  {Marangos}}]{Fleischhauer.2005}%
  \BibitemOpen
  \bibfield  {author} {\bibinfo {author} {\bibfnamefont {M.}~\bibnamefont
  {Fleischhauer}}, \bibinfo {author} {\bibfnamefont {A.}~\bibnamefont
  {Imamoglu}},\ and\ \bibinfo {author} {\bibfnamefont {J.~P.}\ \bibnamefont
  {Marangos}},\ }\href {https://doi.org/\url{10.1103/RevModPhys.77.633}}
  {\bibfield  {journal} {\bibinfo  {journal} {Rev. Mod. Phys.}\ }\textbf
  {\bibinfo {volume} {77}},\ \bibinfo {pages} {633} (\bibinfo {year}
  {2005})}\BibitemShut {NoStop}%
\bibitem [{\citenamefont {Phillips}\ \emph {et~al.}(2001)\citenamefont
  {Phillips}, \citenamefont {Fleischhauer}, \citenamefont {Mair}, \citenamefont
  {Walsworth},\ and\ \citenamefont {Lukin}}]{Phillips.2001}%
  \BibitemOpen
  \bibfield  {author} {\bibinfo {author} {\bibfnamefont {D.~F.}\ \bibnamefont
  {Phillips}}, \bibinfo {author} {\bibfnamefont {A.}~\bibnamefont
  {Fleischhauer}}, \bibinfo {author} {\bibfnamefont {A.}~\bibnamefont {Mair}},
  \bibinfo {author} {\bibfnamefont {R.~L.}\ \bibnamefont {Walsworth}},\ and\
  \bibinfo {author} {\bibfnamefont {M.~D.}\ \bibnamefont {Lukin}},\ }\href
  {https://doi.org/\url{10.1103/PhysRevLett.86.783}} {\bibfield  {journal}
  {\bibinfo  {journal} {Phys. Rev. Lett.}\ }\textbf {\bibinfo {volume} {86}},\
  \bibinfo {pages} {783} (\bibinfo {year} {2001})}\BibitemShut {NoStop}%
\bibitem [{\citenamefont {Tseng}\ \emph {et~al.}(2022)\citenamefont {Tseng},
  \citenamefont {Wei},\ and\ \citenamefont {Chen}}]{Tseng.2022}%
  \BibitemOpen
  \bibfield  {author} {\bibinfo {author} {\bibfnamefont {Y.-C.}\ \bibnamefont
  {Tseng}}, \bibinfo {author} {\bibfnamefont {Y.-C.}\ \bibnamefont {Wei}},\
  and\ \bibinfo {author} {\bibfnamefont {Y.-C.}\ \bibnamefont {Chen}},\ }\href
  {https://doi.org/\url{10.1364/OE.460026}} {\bibfield  {journal} {\bibinfo
  {journal} {{Opt. Express}}\ }\textbf {\bibinfo {volume} {30}},\ \bibinfo
  {pages} {19944} (\bibinfo {year} {2022})}\BibitemShut {NoStop}%
\bibitem [{\citenamefont {Steck}(2019)}]{Steck.2019}%
  \BibitemOpen
  \bibfield  {author} {\bibinfo {author} {\bibfnamefont {D.~A.}\ \bibnamefont
  {Steck}},\ }\href {\url{https://steck.us/alkalidata/cesiumnumbers.pdf}}
  {\bibinfo {title} {Cesium {D} line data}},\ \bibinfo {howpublished}
  {\url{https://steck.us/alkalidata/cesiumnumbers.pdf}} (\bibinfo {year}
  {Revision 2.2.1, 21 November 2019})\BibitemShut {NoStop}%
\bibitem [{\citenamefont {Hsiao}\ \emph {et~al.}(2018)\citenamefont {Hsiao},
  \citenamefont {Tsai}, \citenamefont {Chen}, \citenamefont {Lin},
  \citenamefont {Hung}, \citenamefont {Lee}, \citenamefont {Chen},
  \citenamefont {Chen}, \citenamefont {Yu},\ and\ \citenamefont
  {Chen}}]{Hsiao.2018}%
  \BibitemOpen
  \bibfield  {author} {\bibinfo {author} {\bibfnamefont {Y.-F.}\ \bibnamefont
  {Hsiao}}, \bibinfo {author} {\bibfnamefont {P.-J.}\ \bibnamefont {Tsai}},
  \bibinfo {author} {\bibfnamefont {H.-S.}\ \bibnamefont {Chen}}, \bibinfo
  {author} {\bibfnamefont {S.-X.}\ \bibnamefont {Lin}}, \bibinfo {author}
  {\bibfnamefont {C.-C.}\ \bibnamefont {Hung}}, \bibinfo {author}
  {\bibfnamefont {C.-H.}\ \bibnamefont {Lee}}, \bibinfo {author} {\bibfnamefont
  {Y.-H.}\ \bibnamefont {Chen}}, \bibinfo {author} {\bibfnamefont {Y.-F.}\
  \bibnamefont {Chen}}, \bibinfo {author} {\bibfnamefont {I.~A.}\ \bibnamefont
  {Yu}},\ and\ \bibinfo {author} {\bibfnamefont {Y.-C.}\ \bibnamefont {Chen}},\
  }\href {https://doi.org/\url{10.1103/PhysRevLett.120.183602}} {\bibfield
  {journal} {\bibinfo  {journal} {Phys. Rev. Lett.}\ }\textbf {\bibinfo
  {volume} {120}},\ \bibinfo {pages} {183602} (\bibinfo {year}
  {2018})}\BibitemShut {NoStop}%
\bibitem [{\citenamefont {Gorshkov}\ \emph
  {et~al.}(2007{\natexlab{a}})\citenamefont {Gorshkov}, \citenamefont
  {Andr{\'e}}, \citenamefont {Fleischhauer}, \citenamefont {S{\o}rensen},\ and\
  \citenamefont {Lukin}}]{Gorshkov.2007}%
  \BibitemOpen
  \bibfield  {author} {\bibinfo {author} {\bibfnamefont {A.~V.}\ \bibnamefont
  {Gorshkov}}, \bibinfo {author} {\bibfnamefont {A.}~\bibnamefont {Andr{\'e}}},
  \bibinfo {author} {\bibfnamefont {M.}~\bibnamefont {Fleischhauer}}, \bibinfo
  {author} {\bibfnamefont {A.~S.}\ \bibnamefont {S{\o}rensen}},\ and\ \bibinfo
  {author} {\bibfnamefont {M.~D.}\ \bibnamefont {Lukin}},\ }\href
  {\url{10.1103/PhysRevLett.98.123601}} {\bibfield  {journal} {\bibinfo
  {journal} {Phys. Rev. Lett.}\ }\textbf {\bibinfo {volume} {98}},\ \bibinfo
  {pages} {1} (\bibinfo {year} {2007}{\natexlab{a}})}\BibitemShut {NoStop}%
\bibitem [{\citenamefont {Gorshkov}\ \emph
  {et~al.}(2007{\natexlab{b}})\citenamefont {Gorshkov}, \citenamefont
  {Andr{\'e}}, \citenamefont {Lukin},\ and\ \citenamefont
  {S{\o}rensen}}]{Gorshkov.2007a}%
  \BibitemOpen
  \bibfield  {author} {\bibinfo {author} {\bibfnamefont {A.~V.}\ \bibnamefont
  {Gorshkov}}, \bibinfo {author} {\bibfnamefont {A.}~\bibnamefont {Andr{\'e}}},
  \bibinfo {author} {\bibfnamefont {M.~D.}\ \bibnamefont {Lukin}},\ and\
  \bibinfo {author} {\bibfnamefont {A.~S.}\ \bibnamefont {S{\o}rensen}},\
  }\href {https://doi.org/\url{10.1103/PhysRevA.76.033804}} {\bibfield
  {journal} {\bibinfo  {journal} {Phys. Rev. A}\ }\textbf {\bibinfo {volume}
  {76}},\ \bibinfo {pages} {033804} (\bibinfo {year}
  {2007}{\natexlab{b}})}\BibitemShut {NoStop}%
\bibitem [{\citenamefont {Gorshkov}\ \emph
  {et~al.}(2007{\natexlab{c}})\citenamefont {Gorshkov}, \citenamefont
  {Andr{\'e}}, \citenamefont {Lukin},\ and\ \citenamefont
  {S{\o}rensen}}]{Gorshkov.2007b}%
  \BibitemOpen
  \bibfield  {author} {\bibinfo {author} {\bibfnamefont {A.~V.}\ \bibnamefont
  {Gorshkov}}, \bibinfo {author} {\bibfnamefont {A.}~\bibnamefont {Andr{\'e}}},
  \bibinfo {author} {\bibfnamefont {M.~D.}\ \bibnamefont {Lukin}},\ and\
  \bibinfo {author} {\bibfnamefont {A.~S.}\ \bibnamefont {S{\o}rensen}},\
  }\href {https://doi.org/\url{10.1103/PhysRevA.76.033805}} {\bibfield
  {journal} {\bibinfo  {journal} {Phys. Rev. A}\ }\textbf {\bibinfo {volume}
  {76}},\ \bibinfo {pages} {033805} (\bibinfo {year}
  {2007}{\natexlab{c}})}\BibitemShut {NoStop}%
\bibitem [{\citenamefont {Gorshkov}\ \emph
  {et~al.}(2007{\natexlab{d}})\citenamefont {Gorshkov}, \citenamefont
  {Andr{\'e}}, \citenamefont {Lukin},\ and\ \citenamefont
  {S{\o}rensen}}]{Gorshkov.2007c}%
  \BibitemOpen
  \bibfield  {author} {\bibinfo {author} {\bibfnamefont {A.~V.}\ \bibnamefont
  {Gorshkov}}, \bibinfo {author} {\bibfnamefont {A.}~\bibnamefont {Andr{\'e}}},
  \bibinfo {author} {\bibfnamefont {M.~D.}\ \bibnamefont {Lukin}},\ and\
  \bibinfo {author} {\bibfnamefont {A.~S.}\ \bibnamefont {S{\o}rensen}},\
  }\href {https://doi.org/\url{10.1103/PhysRevA.76.033806}} {\bibfield
  {journal} {\bibinfo  {journal} {Phys. Rev. A}\ }\textbf {\bibinfo {volume}
  {76}},\ \bibinfo {pages} {033806} (\bibinfo {year}
  {2007}{\natexlab{d}})}\BibitemShut {NoStop}%
\bibitem [{\citenamefont {Gorshkov}\ \emph {et~al.}(2008)\citenamefont
  {Gorshkov}, \citenamefont {Calarco}, \citenamefont {Lukin},\ and\
  \citenamefont {S{\o}rensen}}]{Gorshkov.2008d}%
  \BibitemOpen
  \bibfield  {author} {\bibinfo {author} {\bibfnamefont {A.~V.}\ \bibnamefont
  {Gorshkov}}, \bibinfo {author} {\bibfnamefont {T.}~\bibnamefont {Calarco}},
  \bibinfo {author} {\bibfnamefont {M.~D.}\ \bibnamefont {Lukin}},\ and\
  \bibinfo {author} {\bibfnamefont {A.~S.}\ \bibnamefont {S{\o}rensen}},\
  }\href {https://doi.org/\url{10.1103/PhysRevA.77.043806}} {\bibfield
  {journal} {\bibinfo  {journal} {Phys. Rev. A}\ }\textbf {\bibinfo {volume}
  {77}},\ \bibinfo {pages} {043806} (\bibinfo {year} {2008})}\BibitemShut
  {NoStop}%
\bibitem [{\citenamefont {Rakher}\ \emph {et~al.}(2013)\citenamefont {Rakher},
  \citenamefont {Warburton},\ and\ \citenamefont {Treutlein}}]{Rakher.2013}%
  \BibitemOpen
  \bibfield  {author} {\bibinfo {author} {\bibfnamefont {M.~T.}\ \bibnamefont
  {Rakher}}, \bibinfo {author} {\bibfnamefont {R.~J.}\ \bibnamefont
  {Warburton}},\ and\ \bibinfo {author} {\bibfnamefont {P.}~\bibnamefont
  {Treutlein}},\ }\href {https://doi.org/\url{10.1103/PhysRevA.88.053834}}
  {\bibfield  {journal} {\bibinfo  {journal} {Phys. Rev. A}\ }\textbf {\bibinfo
  {volume} {88}},\ \bibinfo {pages} {053834} (\bibinfo {year}
  {2013})}\BibitemShut {NoStop}%
\bibitem [{\citenamefont {Nunn}\ \emph {et~al.}(2013)\citenamefont {Nunn},
  \citenamefont {Langford}, \citenamefont {Kolthammer}, \citenamefont
  {Champion}, \citenamefont {Sprague}, \citenamefont {Michelberger},
  \citenamefont {Jin}, \citenamefont {England},\ and\ \citenamefont
  {Walmsley}}]{Nunn.2013}%
  \BibitemOpen
  \bibfield  {author} {\bibinfo {author} {\bibfnamefont {J.}~\bibnamefont
  {Nunn}}, \bibinfo {author} {\bibfnamefont {N.~K.}\ \bibnamefont {Langford}},
  \bibinfo {author} {\bibfnamefont {W.~S.}\ \bibnamefont {Kolthammer}},
  \bibinfo {author} {\bibfnamefont {T.~F.~M.}\ \bibnamefont {Champion}},
  \bibinfo {author} {\bibfnamefont {M.~R.}\ \bibnamefont {Sprague}}, \bibinfo
  {author} {\bibfnamefont {P.~S.}\ \bibnamefont {Michelberger}}, \bibinfo
  {author} {\bibfnamefont {X.-M.}\ \bibnamefont {Jin}}, \bibinfo {author}
  {\bibfnamefont {D.~G.}\ \bibnamefont {England}},\ and\ \bibinfo {author}
  {\bibfnamefont {I.~A.}\ \bibnamefont {Walmsley}},\ }\href
  {https://doi.org/\url{10.1103/PhysRevLett.110.133601}} {\bibfield  {journal}
  {\bibinfo  {journal} {Phys. Rev. Lett.}\ }\textbf {\bibinfo {volume} {110}},\
  \bibinfo {pages} {133601} (\bibinfo {year} {2013})}\BibitemShut {NoStop}%
\bibitem [{\citenamefont {Guan}\ \emph {et~al.}(2007)\citenamefont {Guan},
  \citenamefont {Chen},\ and\ \citenamefont {Yu}}]{Guan.2007}%
  \BibitemOpen
  \bibfield  {author} {\bibinfo {author} {\bibfnamefont {P.-C.}\ \bibnamefont
  {Guan}}, \bibinfo {author} {\bibfnamefont {Y.-F.}\ \bibnamefont {Chen}},\
  and\ \bibinfo {author} {\bibfnamefont {I.~A.}\ \bibnamefont {Yu}},\ }\href
  {https://doi.org/\url{10.1103/PhysRevA.75.013812}} {\bibfield  {journal}
  {\bibinfo  {journal} {Phys. Rev. A}\ }\textbf {\bibinfo {volume} {75}},\
  \bibinfo {pages} {013812} (\bibinfo {year} {2007})}\BibitemShut {NoStop}%
\bibitem [{\citenamefont {Xu}\ \emph {et~al.}(2013)\citenamefont {Xu},
  \citenamefont {Wu}, \citenamefont {Tian}, \citenamefont {Chen}, \citenamefont
  {Zhang}, \citenamefont {Yan}, \citenamefont {Li}, \citenamefont {Wang},
  \citenamefont {Xie},\ and\ \citenamefont {Peng}}]{Xu.2013}%
  \BibitemOpen
  \bibfield  {author} {\bibinfo {author} {\bibfnamefont {Z.}~\bibnamefont
  {Xu}}, \bibinfo {author} {\bibfnamefont {Y.}~\bibnamefont {Wu}}, \bibinfo
  {author} {\bibfnamefont {L.}~\bibnamefont {Tian}}, \bibinfo {author}
  {\bibfnamefont {L.}~\bibnamefont {Chen}}, \bibinfo {author} {\bibfnamefont
  {Z.}~\bibnamefont {Zhang}}, \bibinfo {author} {\bibfnamefont
  {Z.}~\bibnamefont {Yan}}, \bibinfo {author} {\bibfnamefont {S.}~\bibnamefont
  {Li}}, \bibinfo {author} {\bibfnamefont {H.}~\bibnamefont {Wang}}, \bibinfo
  {author} {\bibfnamefont {C.}~\bibnamefont {Xie}},\ and\ \bibinfo {author}
  {\bibfnamefont {K.}~\bibnamefont {Peng}},\ }\href
  {https://doi.org/\url{10.1103/PhysRevLett.111.240503}} {\bibfield  {journal}
  {\bibinfo  {journal} {Phys. Rev. Lett.}\ }\textbf {\bibinfo {volume} {111}},\
  \bibinfo {pages} {240503} (\bibinfo {year} {2013})}\BibitemShut {NoStop}%
\end{thebibliography}%

\end{document}